\documentclass[acmtog,nonacm,screen]{acmart}
\AtBeginDocument{%
  }

\usepackage{algorithm}
\usepackage{algorithmicx}
\usepackage{algpseudocode}

\DeclareMathOperator*{\Assembly}{\scalebox{1.5}{$\mathbf{A}$}}

\theoremstyle{acmplain}
\newtheorem{remark}{Remark}

\begin{document}

\title{iLogMap: Geodesic Polar Coordinates Parameterization with the Magnetic Laplacian}

\author{Tomás Banduc}
\authornote{All authors contributed equally to this research.}
\email{tbanduc@dim.uchile.cl}
\affiliation{%
  \department{Millennium Institute for Intelligent Healthcare Engineering iHEALTH}
  \institution{Pontificia Universidad Católica de Chile}
  \city{Santiago}
  \country{Chile}
}

\author{Simone Pezzuto}
\authornotemark[1]
\email{simone.pezzuto@unitn.it}
\affiliation{%
  \department{Center for Computational Medicine in Cardiology, Euler Institute}
  \institution{Università della Svizzera italiana}
  \city{Lugano}
  \country{Switzerland}
}
\affiliation{%
  \department{Laboratory of Mathematics for Biology and Medicine, Department of Mathematics}
  \institution{Università di Trento}
  \city{Trento}
  \country{Italy}
}

\author{Francisco Sahli Costabal}
\correspondingauthor
\authornotemark[1]
\email{fsc@ing.puc.cl}
\affiliation{%
  \department{Institute for Biological and Medical Engineering}
  \department{Department of Mechanical and Metallurgical Engineering}
  \department{Millennium Institute for Intelligent Healthcare Engineering iHEALTH}
  \institution{Pontificia Universidad Católica de Chile}
  \city{Santiago}
  \country{Chile}
}
 
\renewcommand{\shortauthors}{Banduc, Pezzuto, and Sahli Costabal}

\begin{teaserfigure}
\centering
\includegraphics[width=\textwidth]{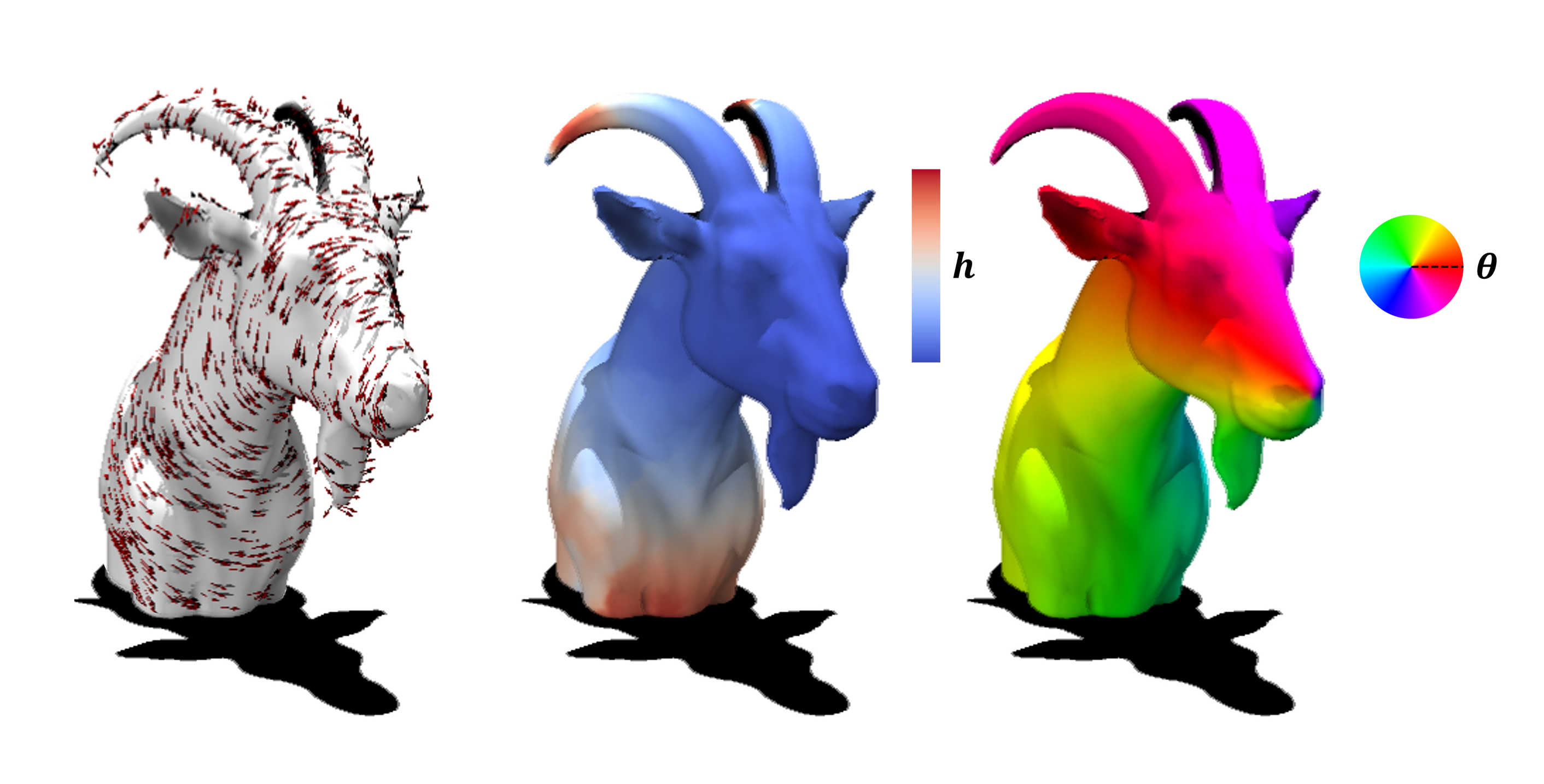}
\caption{Given a base point $\mathrm{p}$ and a geodesic distance solver, iLogMap informs a magnetic Laplacian with a circumferential direction field (left) and a geodesic spread factor (center) to compute the angle component of the logarithmic map (right). This method generalizes to anisotropic metrics and volumetric domains.}
\Description{A goat-shaped surface mesh. Left: circumferential direction induced by a geodesic radial field from a base point located at the goat's muzzle. Center: a scale factor h evaluated across the geometry that increases away from the base point. Right: A geodesic angle map reconstructed with iLogMap using the circumferential direction and scale factor h.}
\end{teaserfigure}

\begin{abstract}
  Geodesic polar coordinates (GPCs) provide an intrinsic parameterization over curved surfaces, but their accurate
  estimation remains challenging, particularly in the presence of anisotropic metrics, high curvature and complex topology. We introduce iLogMap, a method for computing GPCs in curved domains that recasts
  the angular component of the logarithmic map to a ground-state magnetic eigenproblem over the circumferential direction field of geodesic distance. Our method effortlessly extends to anisotropic metric tensors and solid volumes, enabling cylindrical and spherical parameterizations in tetrahedral meshes. Experiments on diverse shapes with varying genus confirm competitive angular accuracy and reduced metric distortion relative to heat-based methods, with improved performance on surfaces with boundary and domains with anisotropy. We demonstrate the utility of iLogMap in computational cardiology applications, where we use it to initialize spiral phases on atrial surfaces and estimate local activation patterns in ventricular models.
\end{abstract}

\begin{CCSXML}
<ccs2012>
   <concept>
       <concept_id>10010147.10010371.10010396</concept_id>
       <concept_desc>Computing methodologies~Shape modeling</concept_desc>
       <concept_significance>500</concept_significance>
       </concept>
   <concept>
       <concept_id>10002950.10003714.10003715.10003750</concept_id>
       <concept_desc>Mathematics of computing~Discretization</concept_desc>
       <concept_significance>300</concept_significance>
       </concept>
   <concept>
       <concept_id>10010405.10010444.10010449</concept_id>
       <concept_desc>Applied computing~Health informatics</concept_desc>
       <concept_significance>100</concept_significance>
       </concept>
 </ccs2012>
\end{CCSXML}

\ccsdesc[500]{Computing methodologies~Shape modeling}
\ccsdesc[300]{Mathematics of computing~Discretization}
\ccsdesc[100]{Applied computing~Health informatics}

\keywords{Logarithmic Map, Magnetic Laplacian, Discrete Differential Geometry, Computational Cardiology}


\maketitle

\section{Introduction}
Local parameterizations of curved surfaces constitute an essential resource in computational science. They facilitate the application of mathematical models defined in non-Euclidean space through the simplified representation of their intrinsic structure. Geodesic polar coordinates (GPCs) are a natural parameterization for this type of domain, extending the classical notion of polar coordinates in flat space to curved surfaces by using the length and direction of the shortest paths emanating from a reference point $\mathrm{p}$.

GPCs are defined via the logarithmic map (log-map), which takes a point $\mathrm{q}$ on the manifold and maps it to a tangent vector $\mathbf{v}$ such that $\mathrm{q}$ is the endpoint reached by traveling along a geodesic starting from $\mathrm{p}$, in the direction of $\mathbf{v}$, for a distance equal to the length of $\mathbf{v}$. The log-map is defined everywhere except on the set of points that can be reached by more than one length-minimizing geodesic (\textit{cut locus}). Geodesic polar coordinates are then obtained by expressing each tangent vector $\mathbf{v}$ in standard polar coordinates $(r,\theta)$.

GPCs have proven useful in several contexts, including geometry processing \cite{sharp2019vector,soliman2025affine}, numerical analysis on manifolds \cite{crane2017,pennec2006} and manifold learning \cite{brun2005,masci2015,malvaer2012}. In cardiac electrophysiology, the log-map has been proposed as a tool for solving the inverse problem of electrocardiography~\cite{grandits2021geasi, grandits2025} and, more recently, for constructing phase fields to initiate spiral wave re-entry on atrial surfaces \cite{banduc2025}.  A recurring challenge in these applications is that existing solvers may produce spurious singularities or lose angular accuracy in the presence of strong anisotropy \cite{banduc2025, jacquemet2012}.

The computation of GPCs on discrete geometries depends directly on how tangent vectors are transported along the surface. The vector heat method \cite{sharp2019vector} exploits the short-time asymptotic behavior of the heat kernel induced by the Levi-Civita connection to approximate parallel transport, which is then used to propagate a radial frame from a base point and obtain a log-map estimate. More recently, the incorporation of infinitesimal translations, in addition to rotations of tangent spaces, has led to the affine heat method \cite{soliman2025affine}. This formulation is based on an augmented connection that generates geodesic polar coordinates with high accuracy, even near the cut locus. Nevertheless, both methods are still sensitive to near-degenerate mesh configurations, do not handle changes of geodesics in response to anisotropy and are not directly applicable to volumes.

In this work, we propose that estimating the angular component of the log-map is equivalent to solving an angular synchronization problem \cite{singer2011}: given a circumferential vector field generated by the geodesic distance from the base point $\mathrm{p}$, one seeks a globally consistent phase assignment $\theta$ that best matches local pairwise orientations. This perspective leads naturally to a spectral relaxation via the \textit{magnetic Laplacian} \cite{fanuel2018,singer2011,xue2024}. The angle field is recovered as the argument of the first non-trivial eigenfunction of a magnetic Laplace operator. The resulting method, which we call \textit{iLogMap}, achieves high angular accuracy, is compatible with different geodesic distance algorithms, supports anisotropic metric tensors and extends to solid volumetric domains.

This manuscript is structured as follows: in section 2 we introduce the methodology for the logarithmic map estimation and describe the numerical approach used to compute GPCs with iLogMap. We also review alternative solvers for comparison, and extend the formulation of our method to anisotropic metrics and volumetric domains. Section 3 reports our experimental results, including convergence and runtime on surfaces with known GPCs, parameterization quality across different shapes, compatibility with other geodesic distance solvers and performance in anisotropic domains and solid volumes. This section concludes with two applications of iLogMap in computational cardiology. Finally, in section 4, we discuss our results, highlight the advantages and limitations of our method, and in section 5 we conclude on the effectiveness of iLogMap as a tool for geometry processing.

\section{Methods}
\subsection{Angular Synchronization Problem}

The \textit{angular synchronization problem} \cite{singer2011,fanuel2018} aims to reconstruct a globally consistent phase field $\phi: M \to [0, 2\pi)$ from local measurements of noisy pairwise angular offsets. We consider a continuous formulation in which the offsets are encoded by a square-integrable vector field $\boldsymbol{\Phi}: M \to TM$ representing the target gradient direction of the phase on a two-dimensional orientable Riemannian manifold $M$ with tangent space $TM$. The goal is to find $\phi$ whose surface gradient $\nabla \phi$ best matches $\boldsymbol{\Phi}$ in the least-squares sense:
\begin{equation}\label{eq:angsynch}
    \min_{\phi:\, M \to [0,2\pi)} \int_M \|\nabla\phi - \boldsymbol{\Phi}\|^2 \, \mathrm{dx}.
\end{equation}
Since $\phi$ takes values in $[0, 2\pi)$, problem~\eqref{eq:angsynch} is equivalent to optimizing over the $SO(2)$ group, and therefore corresponds to a non-convex minimization task.

\subsubsection{Relaxation Via the Magnetic Laplacian} Let $\psi := e^{\iota\phi}$ be the unit-modulus complex lift of the phase. Since $(\nabla - \iota\boldsymbol{\Phi})e^{\iota\phi} = \iota(\nabla\phi - \boldsymbol{\Phi})e^{\iota\phi}$ away from the phase jump region at $2\pi$, a direct computation yields the identity
\begin{equation}\label{eq:key-obs}
    \left\|(\nabla - \iota\boldsymbol{\Phi})\psi\right\|^2 = \|\nabla\phi - \boldsymbol{\Phi}\|^2\quad \text{a.e.}
\end{equation}Therefore, when $\psi = e^{\iota\phi}$, problem~\eqref{eq:angsynch} is equivalent to minimizing $\int_M \|(\nabla - \iota\boldsymbol{\Phi})\psi\|^2 \, \mathrm{d}\mathbf{x}$. Relaxing the pointwise constraint $|\psi| = 1$ to the weaker normalization $\|\psi\|_{L^2(M)} = 1$ 
converts~\eqref{eq:angsynch} into a standard Rayleigh-Ritz minimization problem \cite{Knoppel2013}
\begin{equation}\label{eq:eigenproblem}
\begin{split}
        &\min_{\|\psi\|_{L^2}=1} \int_M \left\|(\nabla - \iota\boldsymbol{\Phi})\psi\right\|^2 \mathrm{dx}\\
    =& \min_{\|\psi\|_{L^2}=1} \langle \psi,\, \mathcal{L}_{\boldsymbol{\Phi}}\,\psi \rangle_{L^2},
\end{split}
\end{equation}
where $\mathcal{L}_{\boldsymbol{\Phi}} := (\nabla - \iota\boldsymbol{\Phi})^2$ is the Laplace operator associated with the magnetic gradient $\nabla - \iota\boldsymbol{\Phi}$ \cite{Baur2025}. The minimizer $\psi^*$ to this problem is the first non-trivial eigenfunction of $\mathcal{L}_{\boldsymbol{\Phi}}$. Then, the recovered phase is $\phi := \arg(\psi^*)$.
\begin{remark}
When $\boldsymbol{\Phi} = \nabla\phi_0$ for some function $\phi_0$, the smallest eigenvalue is zero and the exact phase $e^{\iota\phi_0}$ is recovered. In general, $\psi^*$ provides the best $L^2$ estimate with respect to $\boldsymbol{\Phi}$.
\end{remark}

\subsection{iLogMap on Surfaces}

\begin{figure*}[t]
\centering
\includegraphics[width=0.95\textwidth]{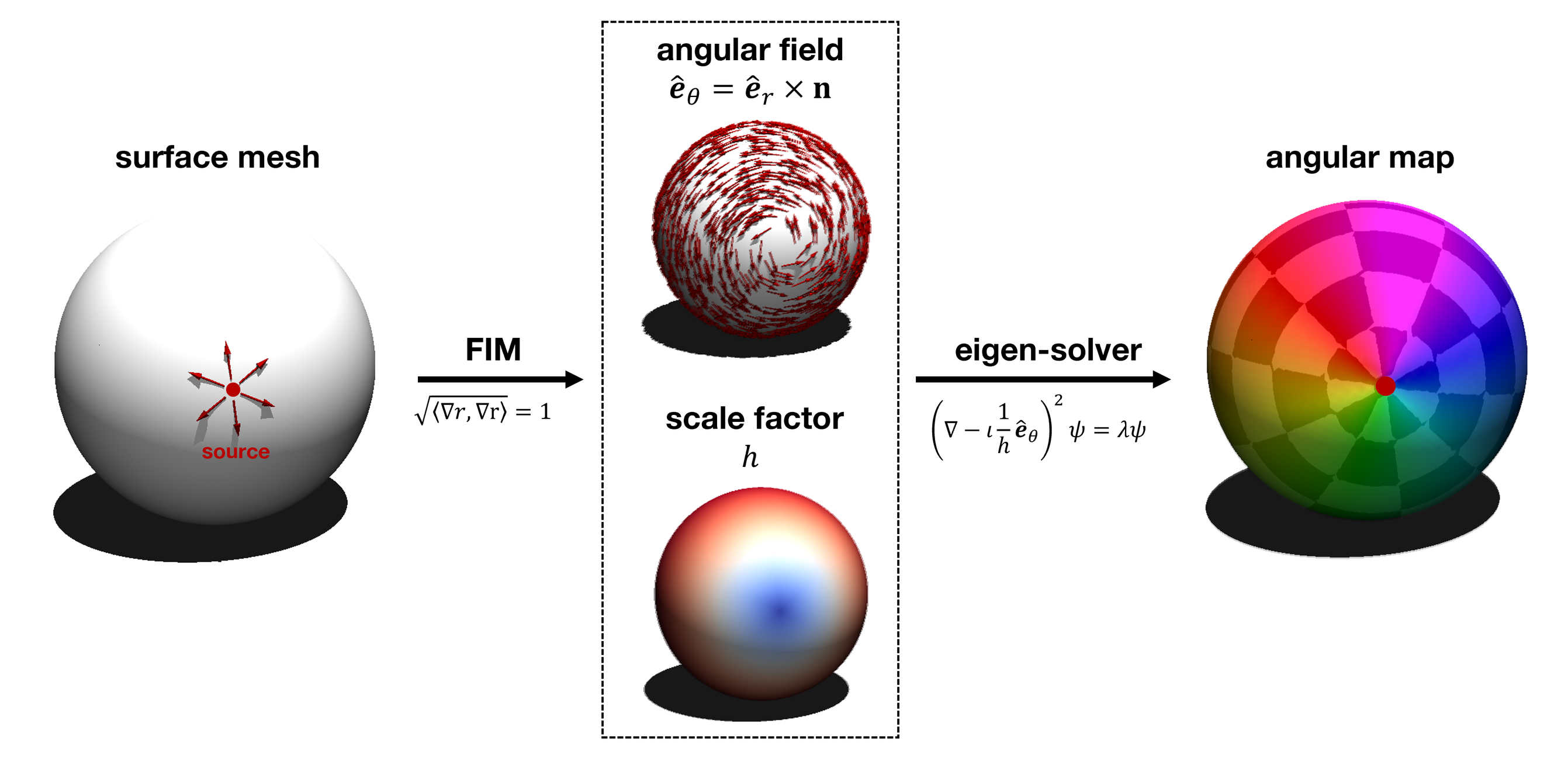}
\caption{iLogMap pipeline. The Fast Iterative Method (FIM) is run from a source point to compute the geodesic distance $r$ and the Jacobi scale factor $h$, which are used to construct the circumferential direction field $\frac{1}{h}\hat{\mathbf{e}}_\theta = \frac{1}{h}(\nabla r \times \mathbf{n})$. Solving the magnetic eigenproblem associated to $\frac{1}{h}\hat{\mathbf{e}}_\theta$  produces the angular map $\theta := \arg(\psi)$.}
\Description{iLogMap schematic: a based point on a curved surface mesh with the eikonal-derived radial direction field emanating from it; the circumferential vector field associated with the radial direction together with a scale Jacobi factor; and the resulting angular map colored with a polar checkerboard texture.}
\label{fig:pipeline}
\end{figure*}

We now apply the angular synchronization framework to recover the angular component of the logarithmic map from a source point $\mathrm{p}$ on a curved surface. The full workflow is illustrated in Figure~\ref{fig:pipeline}.

\subsubsection{Geometric Setup}

Let $r: M \to \mathbb{R}_{\geq 0}$ be the geodesic distance from $\mathrm{p}$ and let $\langle\cdot ,\cdot\rangle$ denote the standard inner product induced by the
Euclidean metric. Since $r$ satisfies the eikonal equation
\[
\sqrt{\langle\nabla r,\nabla r\rangle} = 1,
\]
the direction $\hat{\mathbf{e}}_r := \nabla r$ is a unit vector field pointing radially outward. Along any geodesic level set $\{r = c\}$, the arc-length element $\mathrm{d}s$ and the angular element $\mathrm{d}\theta$ of the log-map are related by
\begin{equation}\label{eq:jacobi-scale}
    \mathrm{d}s = h\,\mathrm{d}\theta,
\end{equation}
where $h: M \to \mathbb{R}$ is the \textit{Jacobi scale factor} encoding how geodesics spread as a function of surface curvature.
\begin{figure}[t]
\centering
\includegraphics[width=0.4\textwidth]{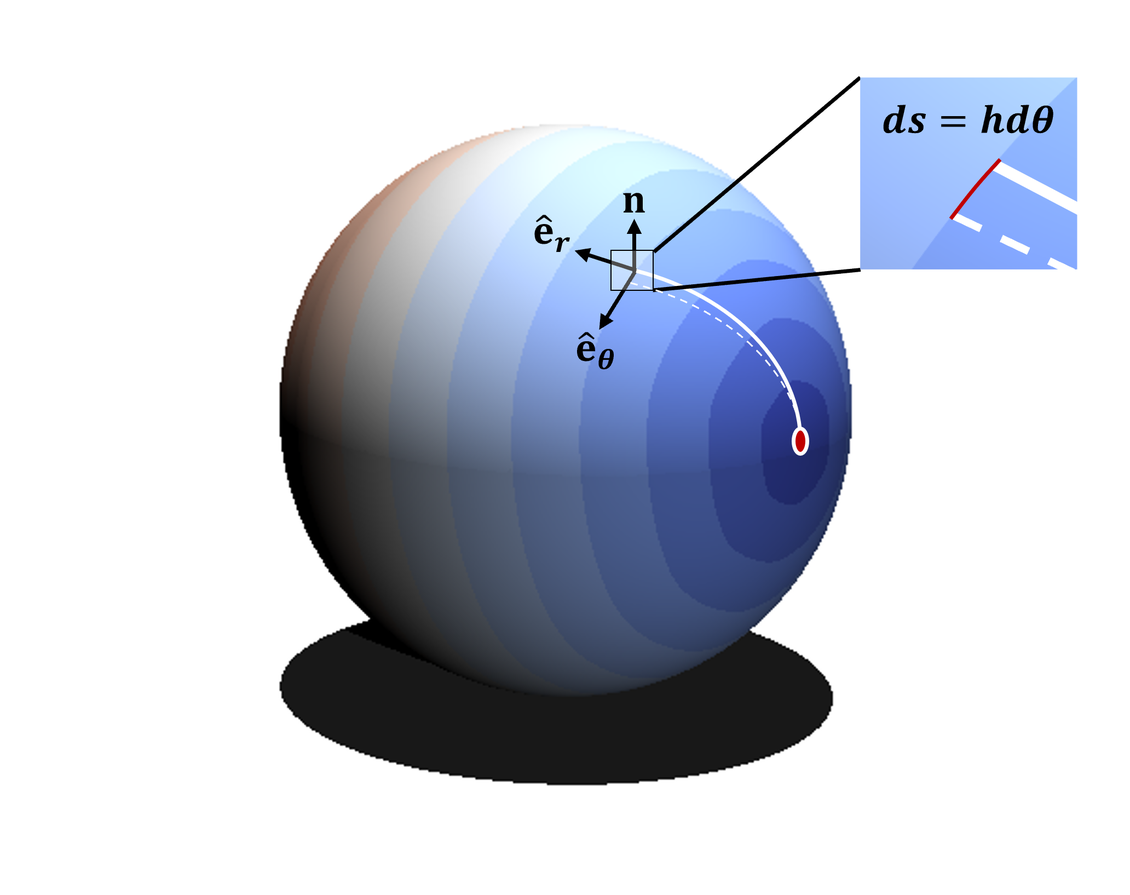}
\caption{Illustration of arc-length element induced by an infinitesimal angular variation of a geodesic on a curved sphere.}
\Description{A short curved arc connecting two nearby geodesics emanating from a common base point on a sphere. The curve shows how infinitesimal changes in arc length relate to small variations in the angle of the logarithmic map on a fixed geodesic level via the Jacobi scale factor.}
\label{fig:arc-length}
\end{figure}
The unit circumferential direction, tangent to the level sets of $r$ and orthogonal to $\hat{\mathbf{e}}_r$, is
\begin{equation}\label{eq:etheta}
    \hat{\mathbf{e}}_\theta := \hat{\mathbf{e}}_r \times \mathbf{n},
\end{equation}
with $\mathbf{n}$ the outward unit surface normal. From~\eqref{eq:jacobi-scale}, the angular component satisfies $\nabla\theta = \hat{\mathbf{e}}_\theta / h$, defining the magnetic vector field for the angular synchronization problem. The angular component is then given by the phase $\theta := \arg(\psi)$ of the function $\psi$ solving
\begin{equation}\label{eq:magnetic-eigen}
    \mathcal{L}_{\theta}\,\psi = \lambda\,\psi,
\end{equation}
to the smallest non-negative eigenvalue $\lambda \geq 0$, with $\mathcal{L}_{\theta}:=(\nabla - \iota \frac{1}{h}\hat{\mathbf{e}}_\theta)^2$.
%We solve this eigenproblem with a shift-and-invert power iteration for $P_1$ Lagrange finite elements.
\subsubsection{The Jacobi Scale Factor} The scalar field $h$ satisfies a scalar Jacobi equation along each geodesic emanating from $\mathrm{p}$ \cite{doCarmo1992}:
\begin{equation}\label{eq:jacobi-transport}
    \left\{
    \begin{array}{l}
        \partial_r^2 h + \kappa\, h = 0, \\[3pt]
        h(\mathrm{p}) = 0, \partial_r h(\mathrm{p}) = 1.
    \end{array}
    \right.
\end{equation}
where $\kappa$ is the Gaussian curvature of $M$. The boundary condition $h(\mathrm{p}) = 0$ expresses that all geodesics propagate from the same location, while the initial condition $\partial_r h(p) = 1$ normalizes the angular spread of geodesic curves. On a flat surface $\kappa(\mathrm{x}) = 0$, and the standard polar coordinate factor $h(r) = r$ is recovered. On a sphere of radius $R$, we get the solution $h(r) = R\sin(r/R)$.

\begin{figure}[t]
\centering
\includegraphics[width=0.45\textwidth]{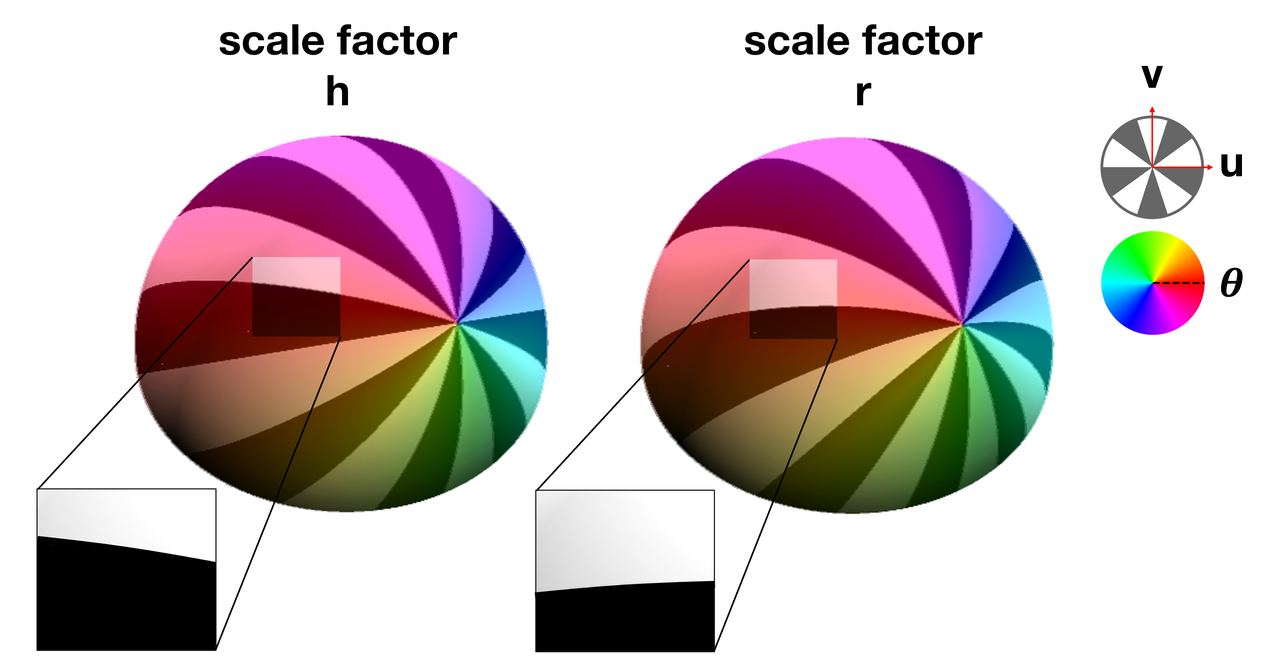}
\caption{Comparison of iLogMap on a half sphere using the solution $h$ to the Jacobi equation \eqref{eq:jacobi-transport} (left) and a radial scale factor $h(r)=r$ (right).}
\Description{Two half-sphere geometries colored with the angular coordinate of the logarithmic map obtained with different scale factors: the left, computed with the Jacobi-equation scale factor, shows evenly spaced angular bands; the right, using the naive scale factor equal to the geodesic distance, shows bands that bend away from the source.}
\label{fig:scale_factor_comparison}
\end{figure}

\begin{remark}
Although we solve for $h$ via the Jacobi equation~\eqref{eq:jacobi-transport}, other scale factors can be used as a substitute estimate of $h$. The simplest choice is $h = r$, which ignores curvature and is exact on flat domains. This may be sufficient for mildly curved surfaces, at the expense of angular accuracy farther from the source point (see Figure~\ref{fig:scale_factor_comparison}).
\end{remark}

\subsubsection{Joint Computation with FIM}

The iLogMap model is compatible with any geodesic distance solver. In our case, we implemented a Fast Iterative Method (FIM) \cite{jeong2008fast, grandits2021} that jointly estimates $h$ at negligible additional cost via forward Euler integration of~\eqref{eq:jacobi-transport} provided the Gaussian curvature of the surface, which we compute using the angle defect formula from \cite{meyer2003}.

\begin{figure*}[t]
\centering
\includegraphics[width=0.9\textwidth]{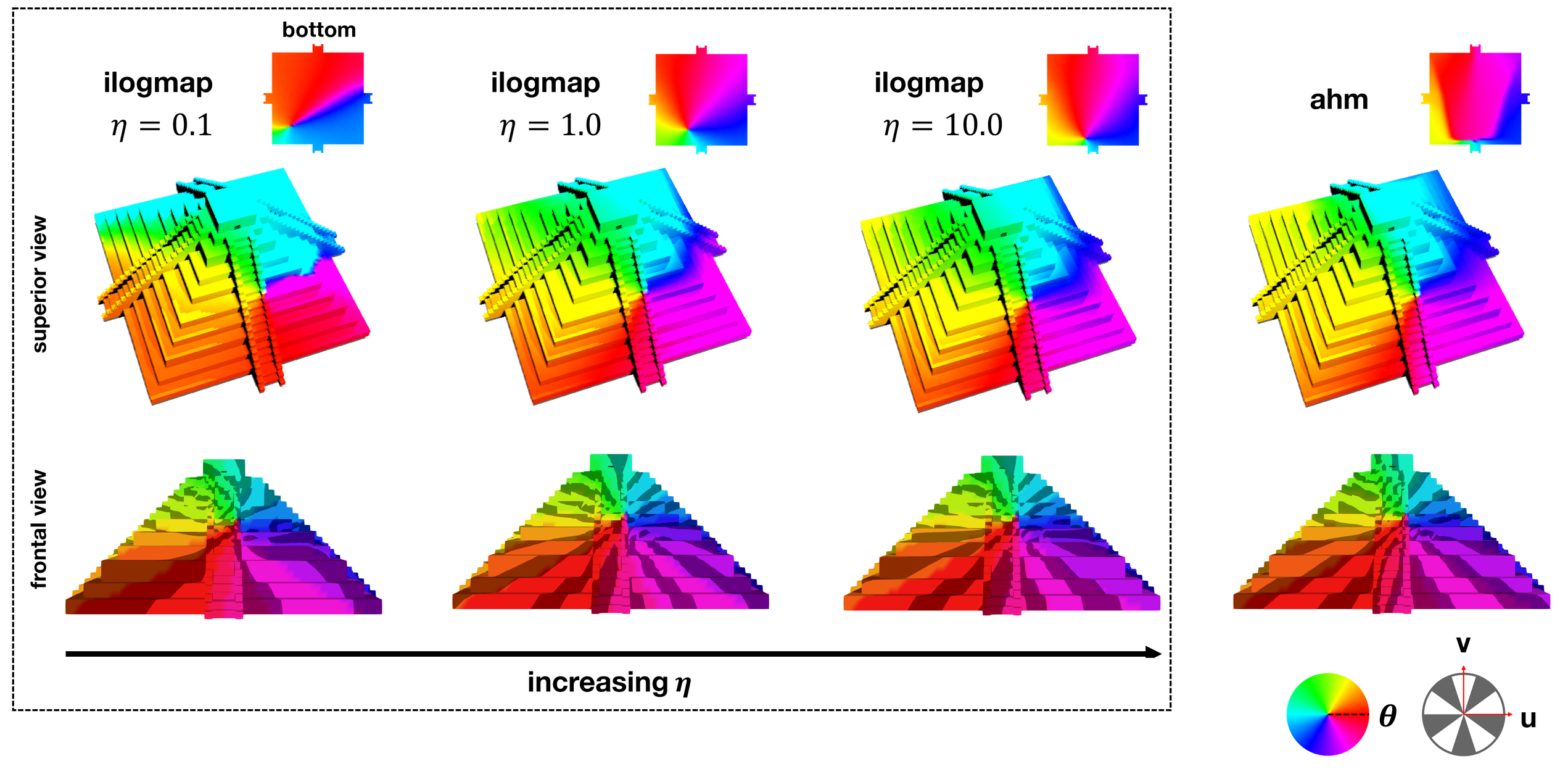}
\caption{Effect of the regularization of Gaussian curvature to iLogMap estimate on a Mayan pyramid mesh (left) and AHM estimate (right, see Section~\ref{sec:ahm}). Superior, inferior and frontal views of a pyramid are shown. Increasing $\eta$ improves robustness near geometric singularities.}%, at the cost of mild angular blurring.}
\Description{Top, bottom and front views of a stepped pyramid mesh, comparing iLogMap's angle output as the smoothing parameter eta is increased from left to right, showing irregular gradients of color for small eta that become smoother for larger eta. In addition to iLogMap, the AHM result is shown for reference.}
\label{fig:epsilon-sensitivity}
\end{figure*}

\begin{remark}
    Gaussian curvature becomes locally undefined on irregular surfaces. This affects the computation of the Jacobi solution $h$ whenever the geometry lacks regularity. To address this issue, we introduce a smoothed curvature field $\tilde{\kappa}$, defined as the solution of the diffusion problem $(\mathrm{id}-\eta\Delta) u=\kappa$. We regard this step as part of pre-processing, since $\kappa$ is an intrinsic quantity independent of the source location $\mathrm{p}$. This regularization enables the method to remain stable on geometries that present sharp corners or steep bending. For sufficiently smooth surfaces, this step is omitted. Figure~\ref{fig:epsilon-sensitivity} illustrates the sensitivity of iLogMap to the parameter $\eta$ on a pyramid mesh with maximum edge size set to $1$.
\end{remark}

%The iLogMap framework supports any geodesic distance solver for the angular synchronization step. We use a Fast Iterative Method (FIM) \cite{TODO:fu2013fim} variant that which solves the eikonal equation by propagating a narrow-band wavefront with an update rule that allows to traverse through both mesh edges and elements. A key advantage of FIM is that we can integrate the computation $h$ can be 

\subsubsection{Compatibility with Alternative Geodesic Methods}
When the geodesic distance solver does not support individual update steps, the field $h$ can instead be obtained by solving an advection--diffusion system with an auxiliary variable $\chi$  encoding the second derivative of $h$:
\begin{equation}\label{eq:jacobi-transport2}
    \left\{
    \begin{array}{l}
        \nabla r \cdot \nabla h - \varepsilon_1\,\Delta h = \chi, \\[3pt]
        \nabla r \cdot \nabla\chi - \varepsilon_2\,\Delta\chi = -\kappa\, h, \\[3pt]
        h(p) = 0,\quad \chi(p) = 1.
    \end{array}
    \right.
\end{equation}

The parameters $\varepsilon_1,\varepsilon_2 > 0$ control the regularized transport of $h$ across geodesics. When these values are small, the solution approximates the true Jacobi field solution. 

System~\eqref{eq:jacobi-transport2} is discretized with standard $P_1$ finite elements and solved using a streamline upwind/Petrov-Galerkin method for the stabilization of streamline directions \cite{BROOKS1982} with $\varepsilon_1=\varepsilon_2$ fixed to a tenth of the maximum edge length in our experiments. 
\subsubsection{iLogMap Algorithm}
We propose two algorithmic versions of our method, which are summarized in Algorithms~\ref{algo:ilogmap} and~\ref{algo:ilogmap-cl}.

The first variant of iLogMap (Algorithm~\ref{algo:ilogmap}) involves solving the eikonal equation, computing the circumferential field estimate of $\theta$ and then solving the magnetic eigenproblem~\eqref{eq:magnetic-eigen}. The output of this procedure corresponds with a complex direction field that smooths angular discontinuities \cite{Knoppel2013} by collapsing them into singularity points inside of the domain. These singularities may distort the log-map parameterization away from the base point~$\mathrm{p}$. For this reason, we introduce a second variant (Algorithm~\ref{algo:ilogmap-cl}) of iLogMap that improves the global accuracy of the GPC output by removing an estimate of the cut locus and performing the angular synchronization step over the resulting cut surface. After the eigenproblem is solved, the complex phase $\psi$ is extended harmonically to the removed region via a Laplace solve and $\theta$ is recovered by taking the argument of the interpolated phase. We use the geodesic-based solver from \cite{mancinelli2021practical} to approximate the cut locus on the input surface.
\begin{figure}[t]
\centering
\includegraphics[width=0.45\textwidth]{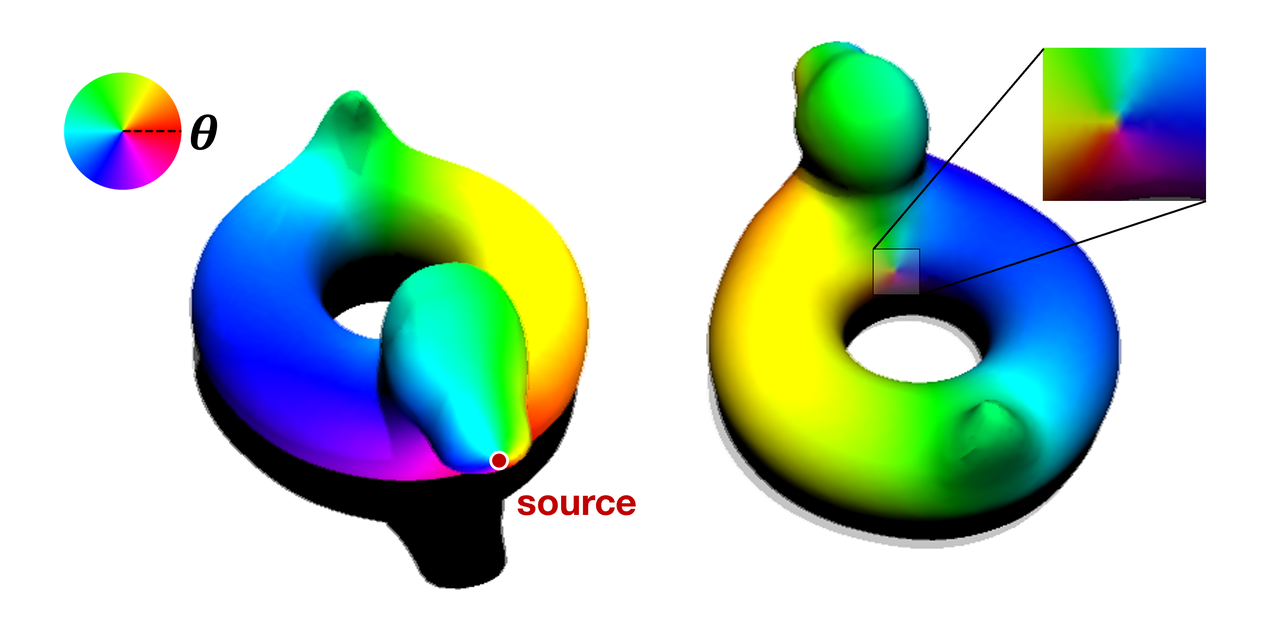}
\caption{Example of iLogMap on a 1-genus surface mesh with collapsed cut locus.}
\Description{A duck-shaped torus mesh colored with the iLogMap angular field theta centered at the beak, showing a globally smooth map with a point singularity at the back of the shape.}
\label{fig:ilogmap-duck}
\end{figure}
\begin{figure}[t]
\centering
\includegraphics[width=0.45\textwidth]{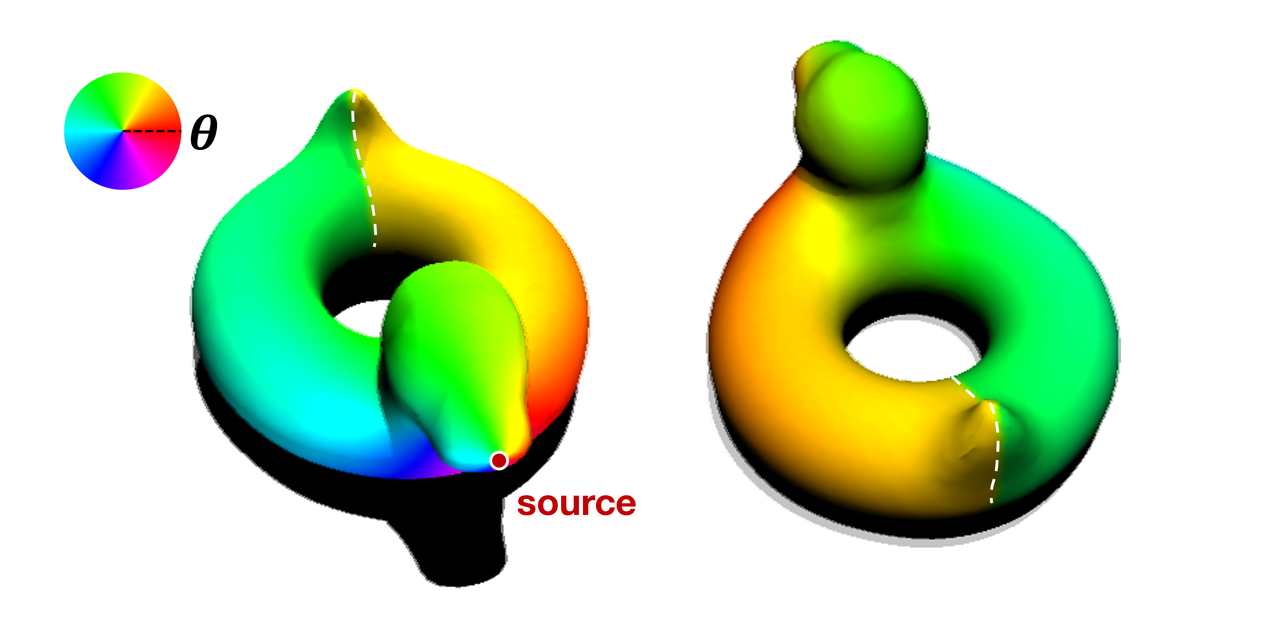}
\caption{Example of iLogMap-cl on a 1-genus surface mesh with preserved cut locus.}
\Description{A duck-shaped torus mesh colored with the iLogMap-cl angular field theta centered at the beak, showing a globally smooth map with a phase jump at the cut locus of the shape.}
\label{fig:ilogmap-cl-duck}
\end{figure}

\begin{algorithm}
\caption{iLogMap}\label{algo:ilogmap}
\textbf{Input:} A source point $\mathrm{p} \in M$ \\
\textbf{Output:} A parameterization $(r,\theta): M \to \mathbb{R}^2$
\begin{algorithmic}[1]
    \State \textbf{Solve} the eikonal equation $\sqrt{\langle\nabla r,\nabla r\rangle} = 1$ from $\mathrm{p}$ (FIM)
    \State \textbf{Estimate} $h$ jointly via~\eqref{eq:jacobi-transport} or via system~\eqref{eq:jacobi-transport2}
    \State \textbf{Compute} the circumferential direction $\hat{\mathbf{e}}_\theta = \nabla r \times \mathbf{n}$
    \State \textbf{Solve} the eigenproblem $\mathcal{L}_{\theta}\,\psi = \lambda\,\psi$
    \State \textbf{Return} $\bigl(r,\;\arg(\psi)\bigr)$
\end{algorithmic}
\end{algorithm}

\begin{algorithm}
\caption{iLogMap with cut locus removal}\label{algo:ilogmap-cl}
\textbf{Input:} A source point $\mathrm{p} \in M$, a detection threshold $t > 0$ \\
\textbf{Output:} A parameterization $(r,\theta): M \to \mathbb{R}^2$
\begin{algorithmic}[1]
    \State \textbf{Solve} the eikonal equation $\sqrt{\langle\nabla r,\nabla r\rangle} = 1$ from $\mathrm{p}$ (FIM)
    \State \textbf{Estimate} $h$ jointly via~\eqref{eq:jacobi-transport} or via system~\eqref{eq:jacobi-transport2}
    \State \textbf{Compute} the circumferential direction $\hat{\mathbf{e}}_\theta = \nabla r \times \mathbf{n}$
    \State \textbf{Detect} the cut locus $\mathcal{C} \subset M$ from $r$ using threshold $t$ \cite{mancinelli2021practical}
    \State \textbf{Solve} the eigenproblem $\mathcal{L}_{\theta}\,\psi = \lambda\,\psi$ on $M \setminus \mathcal{C}$
    \State \textbf{Extend} $\psi$ harmonically to $\mathcal{C}$.
    \State \textbf{Return} $\bigl(r,\;\arg(\psi)\bigr)$
\end{algorithmic}
\end{algorithm}

%\begin{figure*}[t]
%\centering
%\includegraphics[width=0.85\textwidth]{figures-2d-experiments/Diapositiva1.PNG}
%\caption{Angular component $\theta$ of the iLogMap on genus-1, 2, and 3 tori, without (ilog, top) and with (ilog-cl, bottom) cut locus removal. Cut locus removal sharpens the parameterization near regions where multiple geodesics collide, producing cleaner transitions.}
%\label{fig:torus-comparison}
%\end{figure*}

\subsection{Finite Element Discretization}\label{sec:discretization}

%The domain $M$ is discretized as a triangular surface mesh $\mathcal{K} = (V, E, F)$ with $N = |V|$ vertices and $|F|$ triangles. We use $P_1$ Lagrange finite elements; functions that are piecewise linear on each triangle, with one degree of freedom per vertex. In what follows, we describe the local element matrices and the global assembly, which together fully determine the magnetic eigenproblem in a discrete setting.
The surface $M$ is discretized using a triangular mesh $\mathcal{T}$. We use first degree polynomial Lagrange shape functions to construct the eigenproblem~\eqref{eq:magnetic-eigen} in the finite element setting. As such, we obtain a magnetic stiffness matrix $\mathbf{L}_\theta$ defined by
\begin{equation}
\label{eq:local-stiffness}
\mathbf{L}_\theta := \Assembly_{T\in\mathcal{T}} \bigl(\mathbf{K}^{\text{re}}_{\theta,T}+\iota\mathbf{K}^{\text{im}}_{\theta,T}\bigl)
\end{equation}
where $\mathbf{A}$ denotes the finite element assembly operator. $\mathbf{K}_{\theta,T}^{\text{re}}$ and $\mathbf{K}_{\theta,T}^{\text{im}}$ denote, respectively, the symmetric and skew-symmetric parts of the local magnetic stiffness matrix:

\begin{align*}
&\left(\mathbf{K}_{\theta,T}^{\text{re}}\right)_{ij}:=\displaystyle\int_{T}\bigl(\langle\nabla \varphi_i,\nabla\varphi_j\rangle + \varphi_i\varphi_j\langle\mathbf{e}_{h,\theta}, \mathbf{e}_{h,\theta}\rangle\bigl)\mathrm{dx},\\
&\left(\mathbf{K}^{\text{im}}_{\theta,T}\right)_{ij} := \displaystyle\int_{T}\bigl(\varphi_i\langle\mathbf{e}_{h,\theta},\nabla\varphi_j\rangle - \varphi_j\langle\mathbf{e}_{h,\theta},\nabla\varphi_i\rangle\bigr)\mathrm{dx}.
\end{align*}
%Here, the vector field $\mathbf{e}_{h,\theta}$ denotes $\frac{1}{h}\hat{\mathbf{e}}_\theta$ and $G$ is a metric tensor (identity for isotropic conductivity, or a positive-definite tensor for anisotropic domains.   
Here, the vector field $\mathbf{e}_{h,\theta}$ denotes the Jacobi scaled direction $\frac{1}{h}\hat{\mathbf{e}}_\theta$.

The mass matrix for our eigenvalue problem is given in the standard form
\begin{equation}\label{eq:local-mass}
\mathbf{M}:=\Assembly_{T\in\mathcal{T}} \mathbf{M}_T,\quad \left(\mathbf{M}_T\right)_{ij} :=\displaystyle\int_{T}\phi_i\phi_j\mathrm{dx}
\end{equation}
\begin{remark}\label{remark:G}
The FEM formulation~\eqref{eq:local-stiffness} corresponds to a uniformly isotropic metric regime. When directionality and heterogeneity are introduced, the standard inner product $\langle\mathbf{u},\mathbf{v}\rangle:=\mathbf{u}^\top\mathbf{v}$ appearing in the stiffness matrix is replaced by the anisotropic inner product $\langle\mathbf{u},G\mathbf{v}\rangle:=\mathbf{u}^\top G\mathbf{v}$ induced by a symmetric positive-definite metric tensor $G$.
\end{remark}

\subsubsection{Shift-and-Invert Eigensolver} The generalized eigenproblem~\eqref{eq:magnetic-eigen} is formulated as 
\begin{equation}\label{eq:gen-eig}
    \mathbf{L}_\theta\,\boldsymbol{\psi} = \lambda\,\mathbf{M}\,\boldsymbol{\psi}.
\end{equation}
We solve \eqref{eq:gen-eig} with a shift-and-invert power iteration method. Given a small shift $\sigma \geq 0$, we pre-compute a sparse LU factorization of $\mathbf{L}_\theta - \sigma\mathbf{M}$. At each step $k$ with a known iterate  $\psi^{(k)}$, we solve the linear system $(\mathbf{L}_\theta - \sigma\mathbf{M})\psi^{(k+1)} = \mathbf{M}\boldsymbol{\psi}^{(k)}$ and normalize the result:
\begin{equation}\label{eq:power-iter}
    \boldsymbol{\psi}^{(k+1)} \leftarrow \frac{(\mathbf{L}_\theta - \sigma\mathbf{M})^{-1}\mathbf{M}\,\boldsymbol{\psi}^{(k)}}{\bigl\|(\mathbf{L}_\theta - \sigma\mathbf{M})^{-1}\mathbf{M}\,\boldsymbol{\psi}^{(k)}\bigr\|_{\mathbf{M}}}.
\end{equation}
This converges to the eigenfunction with the smallest eigenvalue exceeding $\sigma$. 

Multiple eigenfunctions are obtained sequentially by deflation: before normalizing, each new iterate is projected orthogonally to all previously converged eigenfunctions. In practice, a single eigenfunction suffices for simply connected surfaces. Domains with complex topology (e.g., high-genus tori) may benefit from computing two or more eigenfunctions and selecting the angular field that better adjusts the spread of radial directions.

\begin{remark}
The factorization of $\mathbf{L}_\theta - \sigma\mathbf{M}$ can be reused across all power iterations at no additional cost, so the dominant expense of our algorithm is associated with the LU factorization of the magnetic stiffness matrix, which is comparable to one solve of the affine heat method (see Figure~\ref{fig:convergence}). The factorization must be recomputed whenever the source point changes.
\end{remark}

\subsection{Alternative Methods}

We compare iLogMap against two approaches for computing geodesic polar coordinates. Both are implemented in our experiments and serve as benchmarks for accuracy and robustness.

\subsubsection{Affine Heat Method}\label{sec:ahm}

The Affine Heat Method (AHM) computes a short-time solution to the vector heat equation \cite{soliman2025affine,sharp2019vector} 
\[
    (\mathrm{id}-\tau \mathcal{L})\mathbf{u} = \mathbf{u}_0,
\]
for a small step size $\tau$ on the augmented bundle $TM \oplus \mathbb{R}$. The AHM uses an affine connection Laplacian $\mathcal{L}$ induced by 
\[
\nabla^A :=
\begin{pmatrix}
\nabla & 0 \\
0 & \mathrm{d}  \\
\end{pmatrix} - \begin{pmatrix}
0 & A \\
0 & 0  \\
\end{pmatrix},
\]
In this setting, $\nabla$ is the Levi-Civita connection on $M$, $\mathrm{d}$ is the standard connection on the line bundle $\mathbb{R}$ and $A$ is an Euclidean transformation acting on $M$ represented by a real-valued matrix. On a curved surface, this operator produces the logarithmic map for asymptotic short-time diffusion \cite{soliman2025affine}.

Two variants are introduced in \cite{soliman2025affine} for this method: \textit{localized} AHM, which applies a single global affine connection Laplacian with a trivial translation $A:=\mathrm{id}$; and the \textit{adaptive} AHM, which incorporates a geodesic frame translation \mbox{\(A:=\Phi\circ\mathrm{id}\)}. The adaptive version of the AHM informs the connection with the radial direction, thus incorporating the behavior of geodesics farther from the source. This yields a more accurate estimate of the GPCs, particularly in the vicinity of the cut locus. In the discrete configuration, both methods amount to solving Laplace-like sparse linear systems.

\begin{remark}
Relative to iLogMap, the AHM does not require the Jacobi scale factor $h$ or an explicit eikonal solve, and it produces distance and direction fields simultaneously.   
\end{remark}

\subsubsection{Laplace Interpolator}

As a simpler baseline, we adopt a harmonic extrapolation approach used for the initialization of a complex eikonal solver from cardiac electrophysiology \cite{jacquemet2012,jacquemet2010}. Let $\Gamma$ denote the polygonal boundary of the 1-ring of $\mathrm{p}$ and let $\{\mathbf{v}_i\}_i\subseteq \Gamma$ be its vertices arranged in cyclic order. An angular boundary condition $\alpha_i \in [0, 2\pi)$ is imposed to each vertex $\mathbf{v}_i$ proportionally to its cumulative arc-length around $\Gamma$. That is,
\begin{equation}\label{eq:laplace-bc}
    \alpha_i = 2\pi \cdot \frac{\displaystyle\sum_{k < i}\|\mathbf{v}_{k+1} - \mathbf{v}_k\|}{\displaystyle\sum_k\|\mathbf{v}_{k+1} - \mathbf{v}_k\|}.
\end{equation}
The complex field $\psi: M \to \mathbb{C}$ is then determined by the Dirichlet problem
\begin{equation}\label{eq:laplace-dirichlet}
    -\Delta \psi = 0 \text{ on } M, \qquad \psi = e^{\iota\alpha}\text{ on }\Gamma,
\end{equation}
and the angular component is recovered as $\theta := \arg(\psi)$. Discretized with $P_1$ elements, this reduces to a single real sparse linear system solve, which is the simplest possible approach.

The main limitation of this method is that the boundary condition is assigned purely from the combinatorial arc-length of the 1-ring boundary, without any geodesic distance or curvature information. On flat domains this works well, but on curved surfaces the intrinsic angular spread of geodesics can deviate substantially from the extrinsic 1-ring geometry, leading to degenerate angular fields. % The method also does not incorporate a scale factor, so it cannot account for curvature-induced compression or expansion of geodesics.

\subsection{Anisotropic Domains}\label{sec:aniso}

iLogMap extends naturally to settings where the metric is given by a symmetric positive-definite tensor $G$. The anisotropic geodesic distance satisfies the weighted eikonal equation
\begin{equation}\label{eq:aniso-eikonal}
    \sqrt{\langle\nabla r,G\nabla r\rangle} \;=\; 1.
\end{equation}

A high value along a preferred direction of $G$ shortens the effective geodesic distance in that direction. The FIM eikonal solver accommodates the anisotropic version by redefining the local distance computation using the metric. The FEM solver accommodates the anisotropy by using $G$ in the per-element stiffness assembly~\eqref{eq:local-stiffness}, so no structural change to the algorithm is needed (see Remark~\ref{remark:G}).

Once $r$ is computed, the radial direction $\hat{\mathbf{e}}_r$ is obtained by normalizing the gradient with respect to the induced $G$-norm:
\begin{equation}\label{eq:aniso-er}
    \hat{\mathbf{e}}_r = \frac{\nabla r}{\sqrt{\langle\nabla r,G\nabla r\rangle}}.
\end{equation}
The circumferential direction is then defined as the $G$-cross product $\mathbf{u}\times_G \mathbf{v}:=\sqrt{\det(G)}\cdot G^{-1}(\mathbf{u}\times \mathbf{v}) = \operatorname{cof}(G)\mathbf{u}\times \mathbf{v}$ of $\hat{\mathbf{e}}_r$ and the surface normal $\mathbf{n}$, and is subsequently normalized:
\begin{equation}\label{eq:aniso-etheta}
    \tilde{\mathbf{e}}_\theta = \hat{\mathbf{e}}_r\times_G\mathbf{n},\qquad
    \hat{\mathbf{e}}_\theta = \frac{\tilde{\mathbf{e}}_\theta}{\sqrt{\langle\tilde{\mathbf{e}}_\theta,G\tilde{\mathbf{e}}_\theta\rangle}}.
\end{equation}

This construction ensures that $\hat{\mathbf{e}}_r$ and $\hat{\mathbf{e}}_\theta$ are orthonormal in the tangent plane according to the inner product induced by $G$, mirroring the isotropic case in the metric-distorted geometry.

Likewise, the affine heat method can be adjusted to accommodate anisotropy via intrinsic triangulations~\cite{sharp2019intrinsic}, where edge lengths are adjusted to capture the effects of $G$ over local distances. For each edge $(i,j)$ with unit direction $\hat{\mathbf{d}}_{ij} = (\mathbf{x}_j - \mathbf{x}_i)/\|\mathbf{x}_j - \mathbf{x}_i\|$, the modified edge size is
\begin{equation}\label{eq:intrinsic-edge}
    l^G_{ij} := \frac{\|\mathbf{x}_j - \mathbf{x}_i\|}{\sqrt{\langle\hat{\mathbf{d}}_{ij}\,, G\, \hat{\mathbf{d}}_{ij}\rangle}},
\end{equation}
so that edges aligned with a high-conductivity axis become shorter and edges in low-conductivity directions become longer. The affine heat method is then run on the mesh with these modified edge lengths. We refer to this variant as \textit{AHM-aniso}.
\subsection{iLogMap in Solid Volumes}\label{sec:volumes}

\subsubsection{Volumetric Extension} In a solid tetrahedral mesh, the surface normal $\mathbf{n}$ is absent, so the circumferential direction $\hat{\mathbf{e}}_\theta = (\hat{\mathbf{e}}_r \times \mathbf{n}) / \|\hat{\mathbf{e}}_r \times \mathbf{n}\|$ cannot be directly formed. Instead, an \emph{axial reference direction} $\hat{\mathbf{a}}$ is provided externally to define a consistent stream of vectors from which GPCs can be recovered. The magnetic eigenproblem~\eqref{eq:magnetic-eigen} and the $P_1$ finite element assembly~\eqref{eq:local-stiffness}--\eqref{eq:local-mass} carry over analogously. 

Depending on the behavior of the radial field used to construct the parameterization, we can create cylindrical- or spherical-like GPCs.

\subsubsection{Cylindrical Coordinates} On curved surfaces, the source is determined by a single point $\mathrm{p}$ where the angular state is undefined. In three-dimensional volumes, the singularity extends to a \emph{filament} curve $\mathcal{P}$ threaded through the domain. We exploit this behavior to propagate radial geodesics with boundary condition $r(\mathrm{p})=0$, for $p\in \mathcal{P}$ and construct cylindrical coordinates $(r,\theta,z)$ inside the volume, where the circumferential direction is given by the orthogonal frame completion between the axial and radial directions:
\begin{equation}\label{eq:cyl-etheta}
    \hat{\mathbf{e}}_\theta := \hat{\mathbf{e}}_r \times \hat{\mathbf{e}}_z,\qquad \hat{\mathbf{e}}_z:=\dfrac{\hat{\mathbf{a}}-\langle \hat{\mathbf{e}}_r,\hat{\mathbf{a}}\rangle\hat{\mathbf{e}}_r}{||\hat{\mathbf{a}}-\langle \hat{\mathbf{e}}_r,\hat{\mathbf{a}}\rangle\hat{\mathbf{e}}_r||}.
\end{equation}
% Working here
The computation of the Jacobi factor $h$ is then performed with equation~\eqref{eq:jacobi-transport} for null curvature and velocity $\hat{\mathbf{e}}_r$.
% To solve for $\theta$, the magnetic eigenproblem is solved without any modifications, while 
The component $z$ is obtained through a Poisson solve $\Delta z=\text{div}(\hat{\mathbf{e}}_z)$. % remark: we must assume here that our scalar field is conservative outside of the cut loucs   %the curvature $\kappa$ induced by a least-square fitting $\hat{\mathbf{e}}_z$ 

%EDITING: We justify this procedure as a continuous piling of surfaces in the direction of the filament to assemble the volume.%Neglecting the effects of bending across the shape, the scale factor for the angular direction is set to $h_\theta := r$ and the axial direction is fixed at a constant value $h_z := 1$. Then, we solve two independent magnetic eigenproblems with fields $\frac{1}{r}\hat{\mathbf{e}}_\theta$ and $\hat{\mathbf{e}}_z$  to recover $\theta$ and $z$, respectively.
\begin{figure}[t]
\centering
\includegraphics[width=0.4\textwidth]{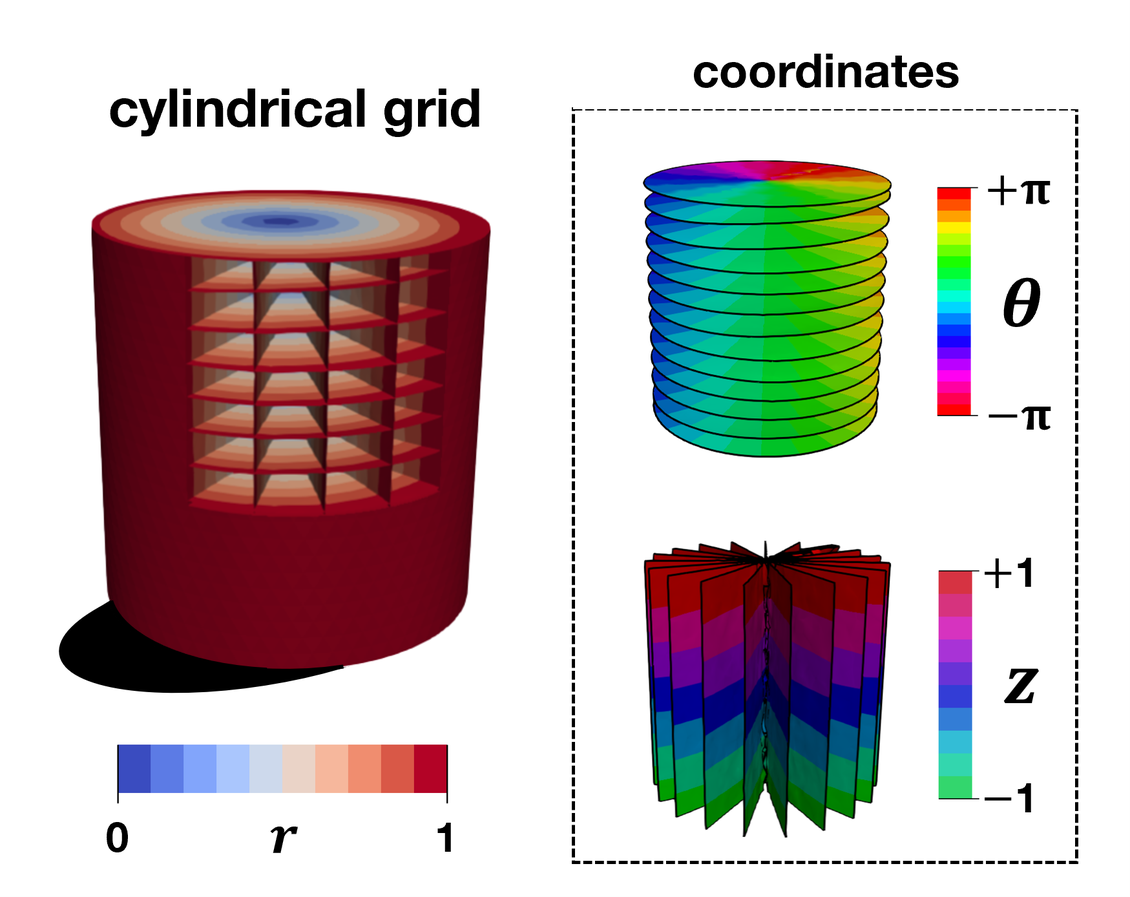}
\caption{Cylindrical coordinates applied to a cylinder volume mesh (layered).
}
\Description{A solid cylindrical mesh with radial and longitudinal layers revealing the interior of the volume; the left, both layers overlapped showing geodesic distance from a central vertical filament; the top right, longitudinal layers showing the angle theta obtained with iLogMap, the bottom right, circumferential layers showing the height of the shape.}
\label{fig:cylinder}
\end{figure}
%\begin{remark}
%EDITING: Mention that the axial direction can take anWe assume that the filament has no component transverse to $\hat{\mathbf{e}}_z$. Under this assumption, the angular coordinate evolves monotonically, preventing phase wrapping and eliminating discontinuities associated with the angle periodicity. Consequently, the recovered phase can be interpreted as an unwrapped angle that is consistent with the shape height up to a global scale factor plus a constant.
%\end{remark}

\subsubsection{Spherical Coordinates} We define the spherical coordinates $(r,\phi,\theta)$ from a single source point $p\in \mathcal{P}$. The azimuthal direction $\hat{\mathbf{e}}_\theta$ is obtained from the cross product of the reference direction and a projected radial field:
\begin{equation}\label{eq:sph-ephi}
    \hat{\mathbf{e}}_\theta := \hat{\mathbf{a}}_r\times\hat{\mathbf{a}} ,\qquad \hat{\mathbf{a}}_r:=\dfrac{\hat{\mathbf{e}}_r-\langle\hat{\mathbf{a}},\hat{\mathbf{e}}_r\rangle\hat{\mathbf{a}}}{\|\hat{\mathbf{e}}_r-\langle\hat{\mathbf{a}},\hat{\mathbf{e}}_r\rangle\hat{\mathbf{a}}\|}.
\end{equation}
The zenith direction is defined as $\hat{\mathbf{e}}_\phi := \hat{\mathbf{e}}_r \times \hat{\mathbf{e}}_\theta$. The scale factors for each angular component are prescribed by solving equation~\eqref{eq:jacobi-transport} with velocity $\hat{\mathbf{e}}_r$ for the zenith scale $h_\phi$ and velocity $\hat{\mathbf{a}}_r$ for the azimuthal scale $h_\theta$.

\begin{figure}[t]
\centering
\includegraphics[width=0.4\textwidth]{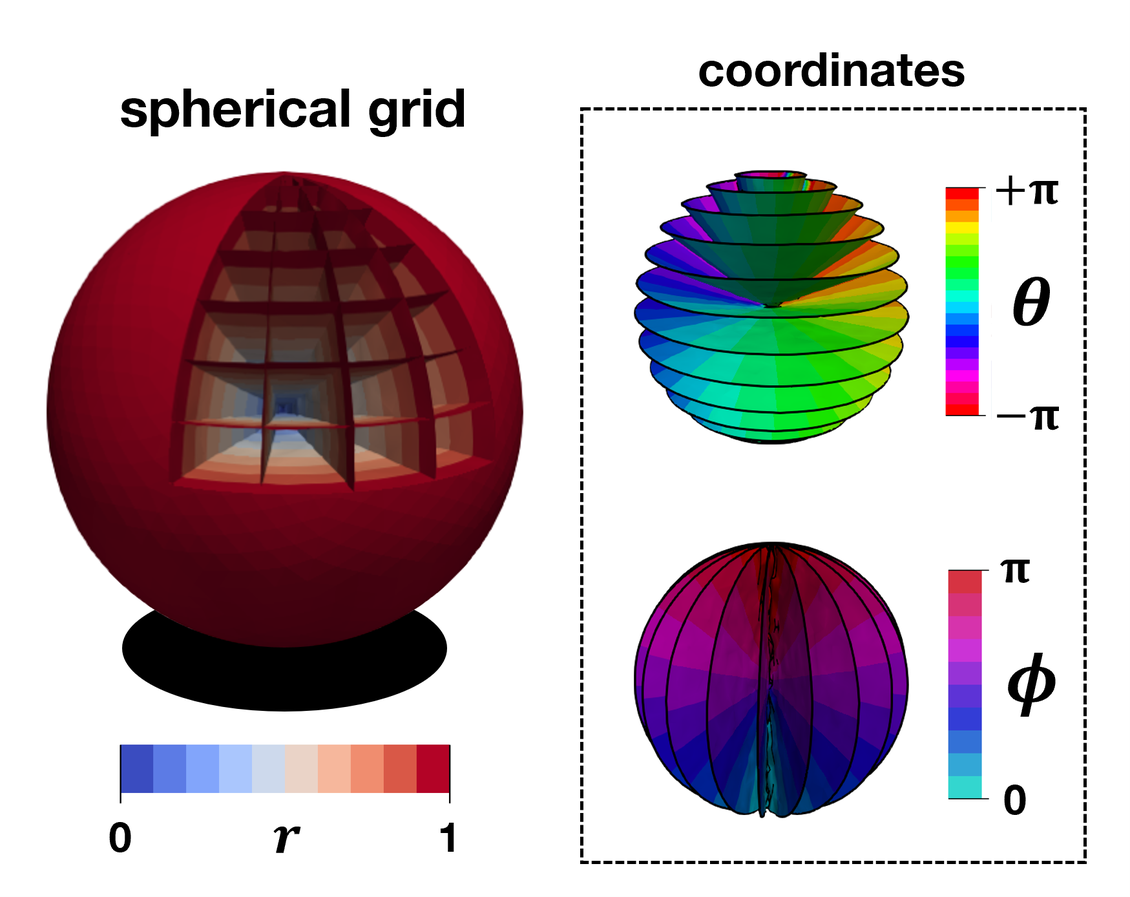}
\caption{Spherical coordinates applied to a sphere volume (layered).}
\Description{A solid spherical mesh with radial and longitudinal layers revealing the interior of the volume; the left, both layers overlapped showing geodesic distance from a central base point; the top right, zenithal layers showing the azimuthal angle theta obtained with iLogMap, the bottom right, circumferential layers showing the zenith output phi across the shape.}
\label{fig:spherical}
\end{figure}
\section{Results}

We evaluate iLogMap on a range of triangular surface meshes of varying genus, resolution and geometric complexity. In all experiments, we compare four methods: iLogMap (without cut locus removal), iLogMap-cl (with cut locus removal; detection threshold $t=\pi/4$), the Affine Heat Method (AHM) \cite{soliman2025affine}, and the Laplace interpolator (\textit{Lap}). All domains are uniformly isotropic unless otherwise stated.

\subsection{Convergence and Runtime}\label{sec:convergence}

\begin{figure*}[t]
\centering
\includegraphics[width=\textwidth]{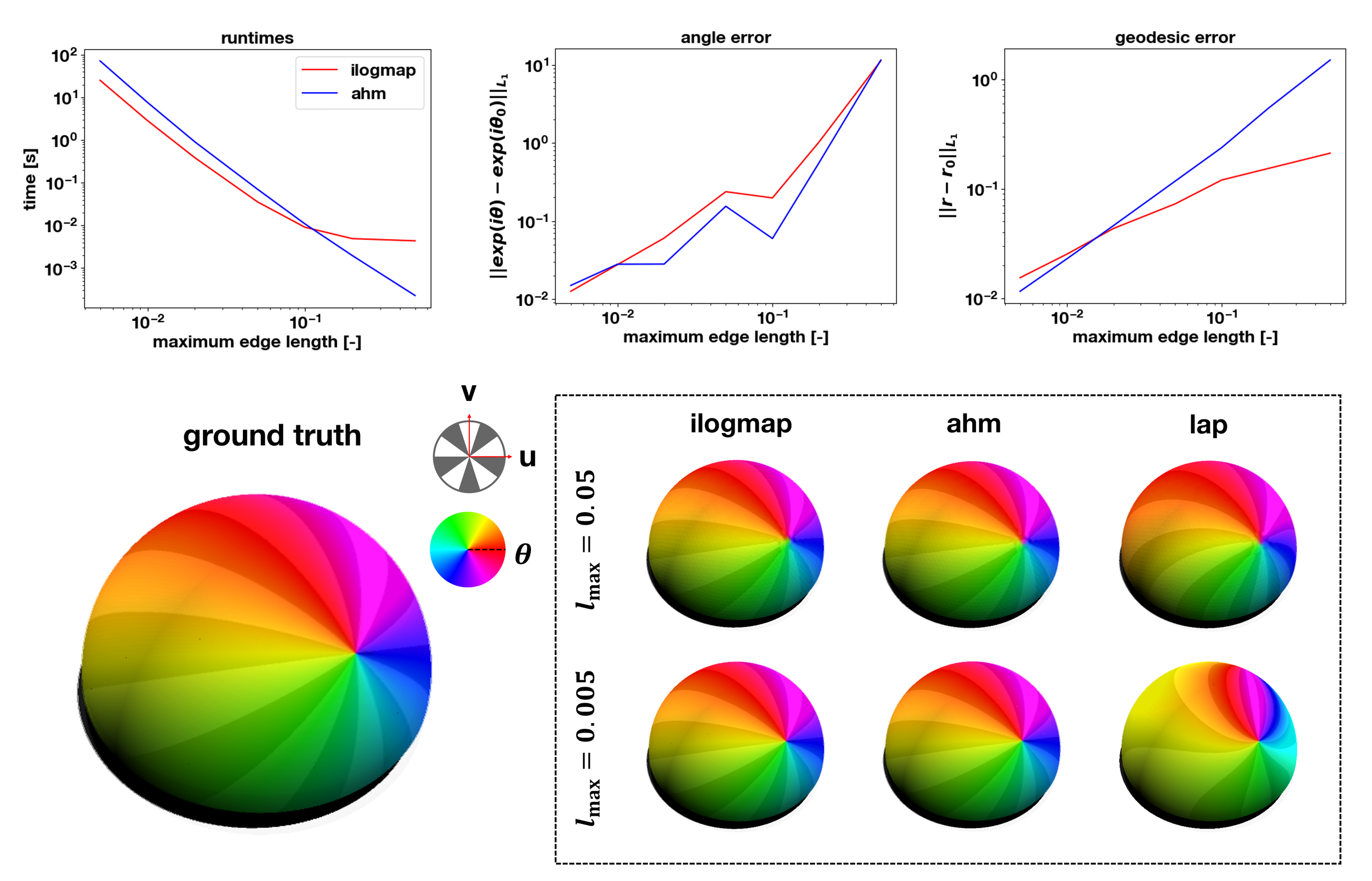}
\caption{Convergence and performance on the half-sphere. Top row: runtime (left), angle error (center) and geodesic distance error (right), as a function of the maximum edge length $l_{\max}$, for iLogMap (blue) and AHM (red). Bottom row: comparison of the three methods at two mesh resolutions ($l_{\max}\in \{0.05, 0.005\}$) alongside the ground truth solution to the logarithmic map angle.}
\Description{Top row: three line plots (runtime, angle error, geodesic distance error), each plotted against maximum mesh edge length for iLogMap and the affine heat method. Runtime trends upward at log-log scale as the mesh resolution increases, while angle and geodesic errors trend downward as the mesh resolution increases. Bottom row: left, ground truth angular field theta on a half sphere with associated radial band texture; right, renderings of iLogMap, the affine heat method and Laplace interpolator on a half-sphere at two mesh resolutions (0.05 and 0.005), with iLogMap and AHM visually closer to the ground truth, and Lap presenting visible bending.}
\label{fig:convergence}
\end{figure*}

Figure~\ref{fig:convergence} reports quantitative convergence of iLogMap and AHM on a sequence of half-sphere meshes with decreasing maximum edge length $l_{\max}$. On the half-sphere, the ground truth log-map is available analytically. The radial component is the geodesic distance $r(\mathrm{x}) := \arccos\left(\mathrm{p}^\top\mathrm{x}\right)$ and the angular component is the azimuthal angle of the geodesic direction. Both iLogMap and AHM exhibit convergent behavior as $l_{\max} \to 0$ in both angle error and geodesic distance error.

The execution times of a C++ implementation of iLogMap bound to Python were registered for a single-core process on an Apple
M2 Ultra machine (24-core CPU and 64 GB unified memory). For the AHM, we used the vector heat solver from the \texttt{potpourri3d} library. The runtime of iLogMap (40[ms], $l_{\max}$=0.05; 25[s], $l_{\max}$=0.005) scales comparably to that of the affine heat method (70[ms],  $l_{\max}$=0.05; 70[s], $l_{\max}$=0.005), with iLogMap yielding higher computational cost at low resolution and competitive execution times as mesh quality increases. The comparison at high-resolution levels between the iLogMap, AHM and Lap suggests that the first two methods produce smooth, concentric angular fields, while the last creates visible distortion that exacerbates with mesh resolution.

\subsection{Angular Accuracy on Tori}

\begin{figure*}[t]
\centering
\includegraphics[width=\textwidth]{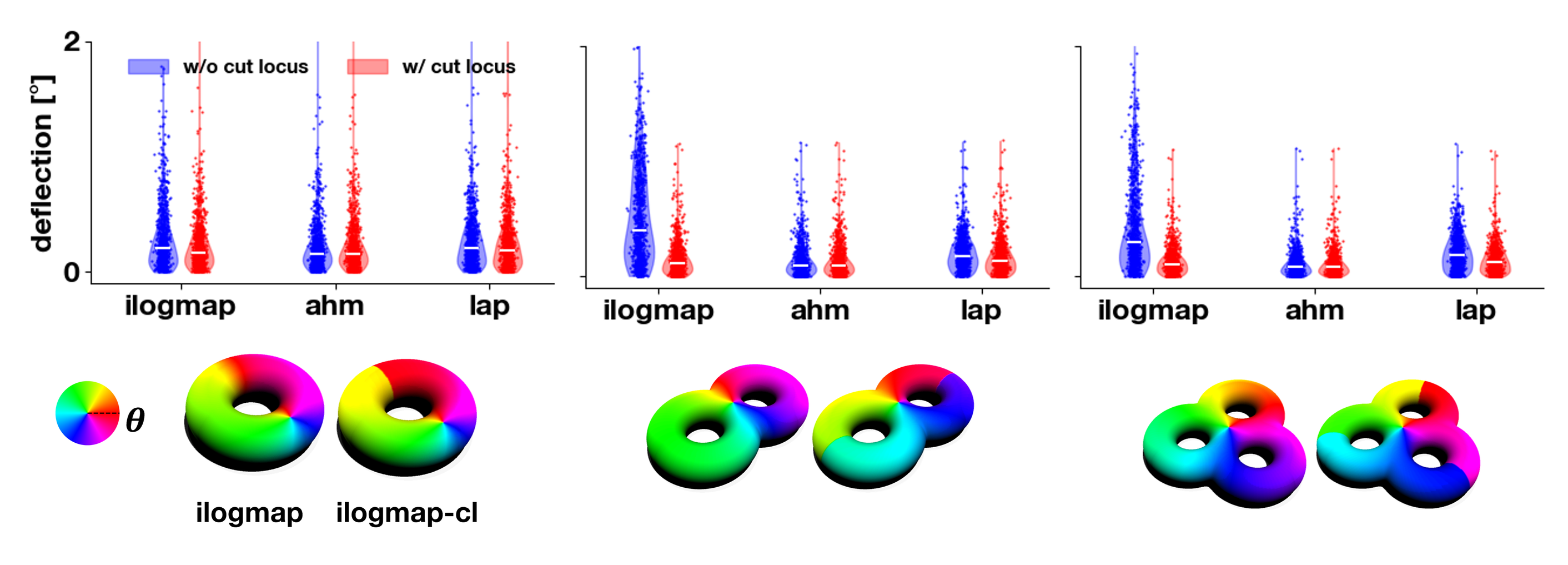}
\caption{Angle deflection along geodesic paths on genus 1, 2, and 3 tori. Violin plots compare ilogmap, ahm and lap with (red) and without (blue) cut locus removal. Lower values indicate that the angular field varies less along geodesic rays, corresponding to a more accurate log-map.
% FSC, use blue and red for the 2 methods, like in the previous figure. Also, what are the units?
}
\Description{Top row: Violin plots showing the distribution of angle deflection along geodesic paths for iLogMap, the affine heat method and the Laplace interpolator, each with and without cut locus removal, grouped by genus-1, genus-2 and genus-3 tori. Bottom row: 1-torus, 2-torus and 3-torus colored with iLogMap and iLogMap-cl outputs from a fixed base point.}
\label{fig:tori-accuracy}
\end{figure*}

We quantify angular accuracy through the angle deflection of randomly chosen geodesic paths emanating from uniformly distributed locations. We compute the angle deflection as the mean angle difference between consecutive nodes in a geodesic path. A perfectly accurate log-map assigns constant $\theta$ along each geodesic, so lower deflection values indicate higher accuracy.

In Figure~\ref{fig:tori-accuracy} we report angle deflection tori of genus 1, 2, and 3 from a family of $10$ base points, each sampled with $100$ geodesic paths of lengths ranging from 10 to 20\% the maximum radial distance. We assess all methods, with and without cut locus removal. For iLogMap and Lap, the removal of the cut locus reduces deflection outcomes, while AHM predictions remain unaffected. For the cut locus removal case, the Laplacian interpolator (median deflection of 0.19° on 1-torus; 0.14° on 2-torus; 0.12° on 3-torus) and iLogMap (median 0.17° on 1-torus; 0.12° on 2-torus; 0.11° on 3-torus) present similar performance, while the AHM persistently achieves the lowest deflection (median 0.16° on 1-torus; 0.10° on 2-torus; 0.08° on 3-torus). Without the cut locus removal, iLogMap presents slightly increased median local deflection compared to the other methods, presenting 0.21 to 0.41 degrees of error, while Lap and AHM register deflection values ranging from 0.17° to 0.20° and 0.08° to 0.16°, respectively.

%On the 1-torus, all approaches show similar trends of angle deflection with median values ranging from X to Y, having the AHM slightly improved accuracy with respe. The median of the AHM remains unchanged to the cut locus removal (), while iLogMap and Lap present noticeable improvement shows heavier tails than ahm and lap, reflecting phase distortion introduced by the cut locus entering the domain. With cut locus removal, the median deflection of ilog-cl drops substantially and the distribution tightens, becoming competitive with ahm. On genus-2 and genus-3 surfaces, all methods achieve broadly similar median deflection, with cut locus removal consistently reducing the variance across methods. The visual angular fields on the same meshes are shown in at the bottom of the figure, where ilog-cl produces clean cuts at the fold regions compared to ilog, which creates smooth transitions instead of phase discontinuities.

%\begin{figure*}[t]
%\centering
%\includegraphics[width=\textwidth]{figures-2d-experiments/Diapositiva3.PNG}
%\caption{Comparison of constructed angular field $\theta$ on the goat mesh. Top row: angular field $\theta$ (hue-coded). Bottom row: metric distortion $\delta = \max(1/\sigma_1,\sigma_2)$, where $\sigma_1, \sigma_2$ are the singular values of the GPC parameterization Jacobian per element. Values close to $1$ indicate a local isometry.}
%\label{fig:goat-comparison}
%\end{figure*}

\begin{figure*}[t]
\centering
\includegraphics[width=\textwidth]{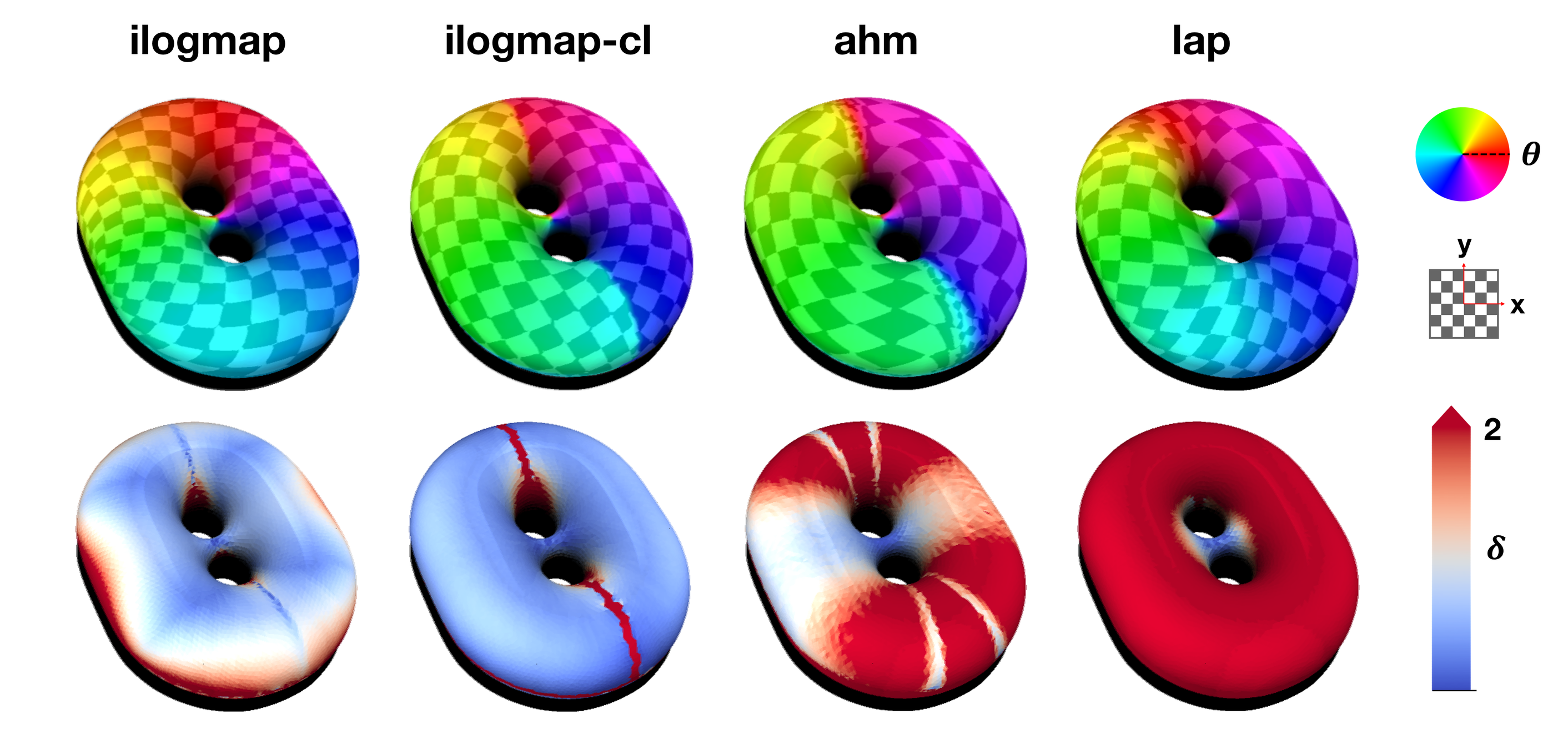}
\caption{Constructed angular fields (top) and polar checkerboard textures (bottom) for all methods on a double torus with a constricted central neck. Uniform spacing of the checkerboard pattern indicates a low-distortion parameterization.}
\Description{Renderings of iLogMap, iLogMap-cl, the affine heat method and the Laplace interpolator over a 2-torus with a pinched central neck, showing the angular field theta, a polar checkerboard texture and metric distortion delta; the distortion from iLogMap-cl remains predominantly close to 1, and the checkerboard pattern stays evenly spaced across the neck and toward the cut locus of the shape; the other methods show visible distortion near the cut locus, with AHM and Lap having the most noticeable distortion.}
\label{fig:torus-texture}
\end{figure*}

On Figure~\ref{fig:torus-texture} we compare the angular estimate $\theta$ and the metric distortion $\delta$ across all four methods on a double torus with a constricted central neck where a base point is placed. The metric distortion $\delta$ is defined as the maximum deviation from unit scaling:
\begin{align}
    \delta:=\max(1/\sigma_1,\sigma_2),
\end{align}
with $\sigma_1\leq \sigma_2$ the singular values of the surface gradient of the map 
\begin{align*}
\mathrm{x}\mapsto\bigl(r(\mathrm{x})\cos(\theta(\mathrm{x})),r(\mathrm{x})\sin(\theta(\mathrm{x}))\bigl).
\end{align*}

We observe that all methods produce a smooth global mapping away from the cut locus and exhibit nearly isometric behavior in the vicinity of the base point. However, substantial differences emerge farther from the bottleneck region. Among all approaches, iLogMap-cl yields the angular approximation with the lowest distortion and exhibits a highly uniform cartesian structure, as evidenced by the induced cartesian checkerboard pattern. In contrast, iLogMap shows mild distortion near the cut locus area. Both approaches are followed by AHM and Lap, with AHM showing an extended patch of metric distortion where the cut locus manifests, and Lap producing the largest metric distortion across most of the domain, except within the central neck of the torus.

\subsection{Surfaces with a Boundary}
In domains with a boundary, vector heat methods typically employ homogeneous Neumann boundary conditions for geodesic distance computation. However, this constraint introduces boundary artifacts arising from reflected contributions of the short-time heat kernel, which become significant within the boundary layer \cite{crane2013heat,avramidi1995}. These effects perturb the gradient field from which the parameterization is recovered.

Since our method is formulated independently of heat diffusion, it can incorporate the no-flux constraint without the boundary effects associated with diffusion asymptotics. This allows our method to accurately estimate GPCs even in domains with extended boundary regions.
\begin{figure}[t]
\centering
\includegraphics[width=0.35\textwidth]{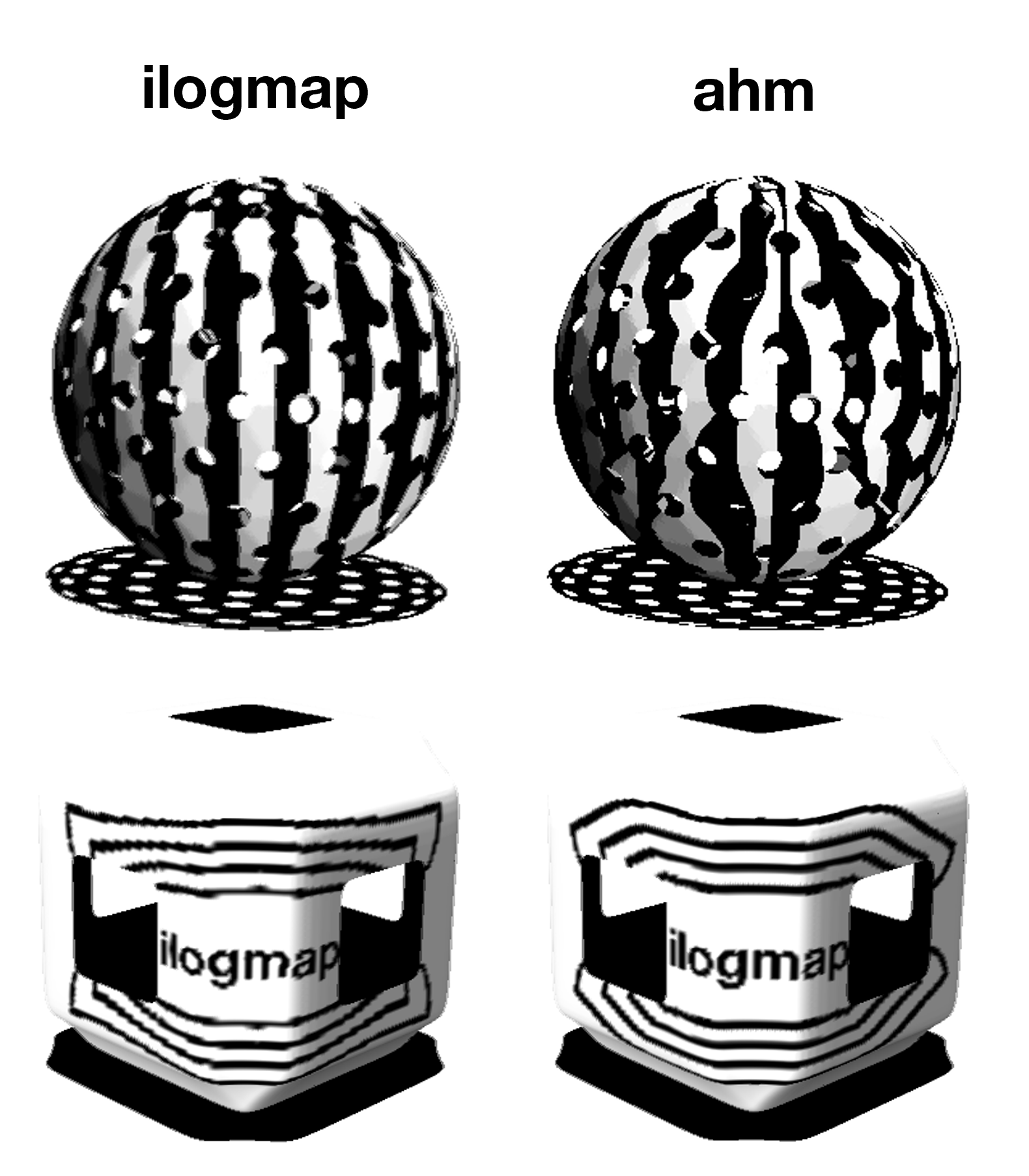}
\caption{Application of textures on perforated surfaces with iLogMap (left) and AHM (right). Top: Black and white stripes on a sphere with circular perforations. Bottom: Nested rectangles and projected text ``ilogmap'' on a beveled cube with squared perforations.}
\Description{Top: a sphere with circular perforations textured with black-and-white stripes, comparing iLogMap (undistorted stripes) and the affine heat method (warped stripes). Bottom: a beveled cube with rectangular perforations textured with nested rectangles and the text ``ilogmap'', comparing iLogMap (undistorted) and the affine heat method (warped near one of the perforation boundaries).}
\label{fig:textures}
\end{figure}
On the example of a perforated sphere in Figure~\ref{fig:textures}, iLogMap produces smooth, undistorted stripe patterns. However, AHM carries significant warping of the added texture, confirming that the parameterization is unable to mitigate the effect of heat reflection at the boundary. In the same Figure, we show a perforated beveled cube with nested rectangles and the text ``ilogmap''. Although both iLogMap and AHM place the texture without visible seams or shearing, the AHM seems to fold the image around one of the perforations.
%This application confirms that iLogMap produces parameterizations of sufficient quality for practical texture mapping tasks, and that the resulting UV coordinates are compatible with standard rendering pipelines.
\subsection{Framework Compatibility}
\begin{figure*}[t]
\centering
\includegraphics[width=0.95\textwidth]{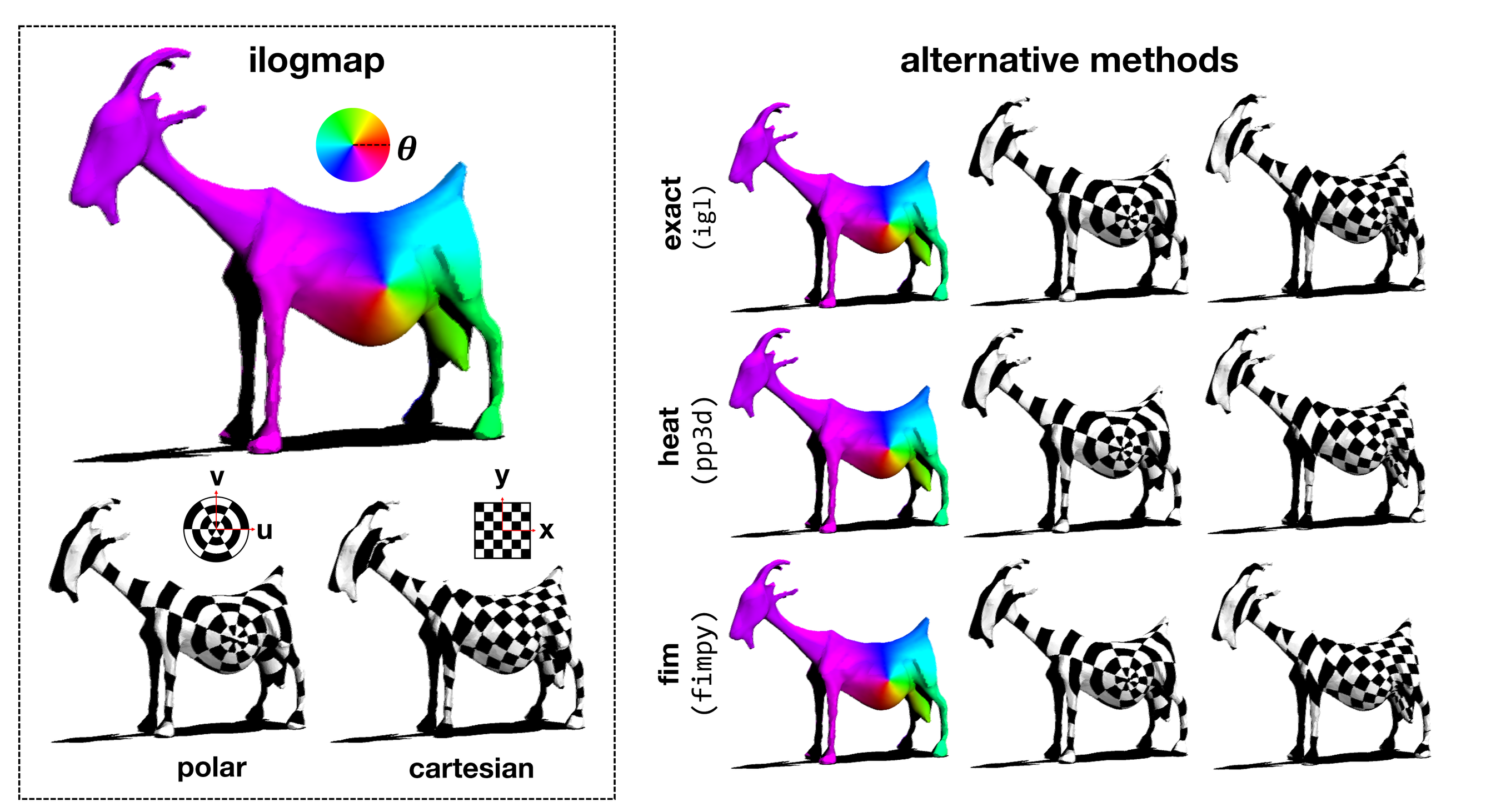}
\caption{iLogMap on the goat mesh for different geodesic solvers. Left panel, visualized polar angle $\theta$ (top) with polar (bottom left, $u$-$v$ checkerboard) and cartesian (bottom right, $x$-$y$ checkerboard) coordinate patterns. Right panel, comparison against three alternative distance-based solvers: exact geodesic (exact/igl), the heat method (heat/pp3d), and FIM (fim/fimpy).}
\Description{A goat shaped mesh with angular map coloring, cartesian and polar checkerboard texture produced by iLogMap, repeated for three different geodesic-distance solvers (exact geodesic solver, the heat method, and a standalone FIM solver), with visually consistent angular fields and checkerboard spacing across all methods.}
\label{fig:goat-compat}
\end{figure*}

Since the angular synchronization step is decoupled from the specific distance computation, we evaluate our solver for three different geodesic distance backends: the exact geodesic algorithm (taken from \texttt{libigl}), the heat method (taken from \texttt{potpourri3d}) and a standalone FIM implementation (taken from \texttt{fim-python}). 

Figure~\ref{fig:goat-compat} shows the iLogMap parameterization on a goat mesh with polar and cartesian visualization modes for the different solvers. Across all cases, the constructed phase remains unchanged, presents angular consistency for the construction of checkerboard textures and displays spatial regularity near the source.
\subsection{Base Point Shifting}
To assess the accuracy of the domain flattening induced by our GPC parameterization, we construct the radial and angular components from a fixed source point, and then use it to compute geodesic distances from points other than the source by calculating Euclidean distances in the GPCs.

\begin{figure*}[t]
\centering
\includegraphics[width=\textwidth]{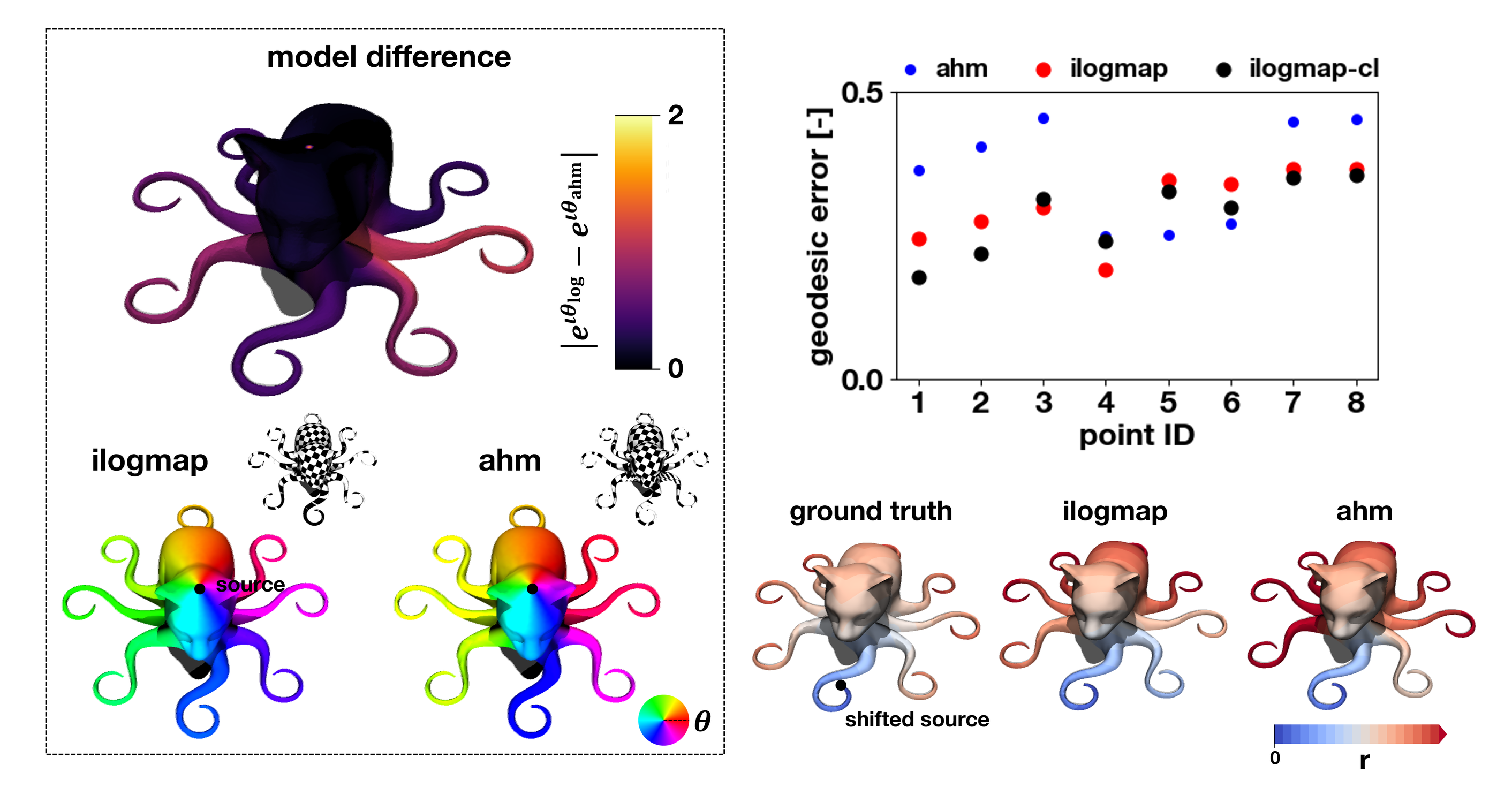}
\caption{Comparison of iLogMap, iLogMap-cl and AHM on the catopus model. Left: pointwise angular difference $|e^{i\theta_{\mathrm{ilog}}} - e^{i\theta_{\mathrm{ahm}}}|$ between iLogMap and AHM (top), with angular fields and cartesian checkerboard pattern (bottom). Right: geodesic distance comparison scatter plot for multiple source points (top) and visualized radial distance field $r$ for a shifted source (bottom).
}
\Description{Left: top, a cat-shaped model with eight tentacles colored by the pointwise angular difference between iLogMap and the affine heat method, small over the central body of the shape, where the base point is located, and larger toward the tentacles; bottom, the angular fields and cartesian checkerboard textures produced by each method. Right: top, a scatter plot comparing radial distances recomputed from shifted source points on each tentacle tip against exact geodesic distances between tentacle tips, with iLogMap and iLogMap-cl points lying closer to ground-truth values than the affine heat method; bottom, example of geodesic distance from a tentacle tip and recomputed distance with iLogMap and AHM.}
\label{fig:catopus}
\end{figure*}

We apply this procedure to a catopus shape in Figure~\ref{fig:catopus}. Specifically, for each tentacle tip, we compute the shifted radial distances and evaluate their accuracy by measuring the mean relative difference with respect to the exact geodesic distances to the remaining tips. The difference between the angular maps (left) of iLogMap and AHM remains predominantly small over the central body of the model, indicating strong agreement between the two methods. The largest discrepancies are concentrated along the tentacles, where the effects of curvature are more pronounced. For the shifted radial distance estimates, iLogMap consistently yields lower errors than AHM, with errors ranging from 19\% to 37\%. Similar results are obtained after the removal of the cut locus, with iLogMap-cl showing a slight reduction in error compared to iLogMap in some cases. %indicating a more accurate intrinsic distance estimate.

\subsection{Gallery of Diverse Shapes}
\begin{figure*}[t]
\centering
\includegraphics[width=\textwidth]{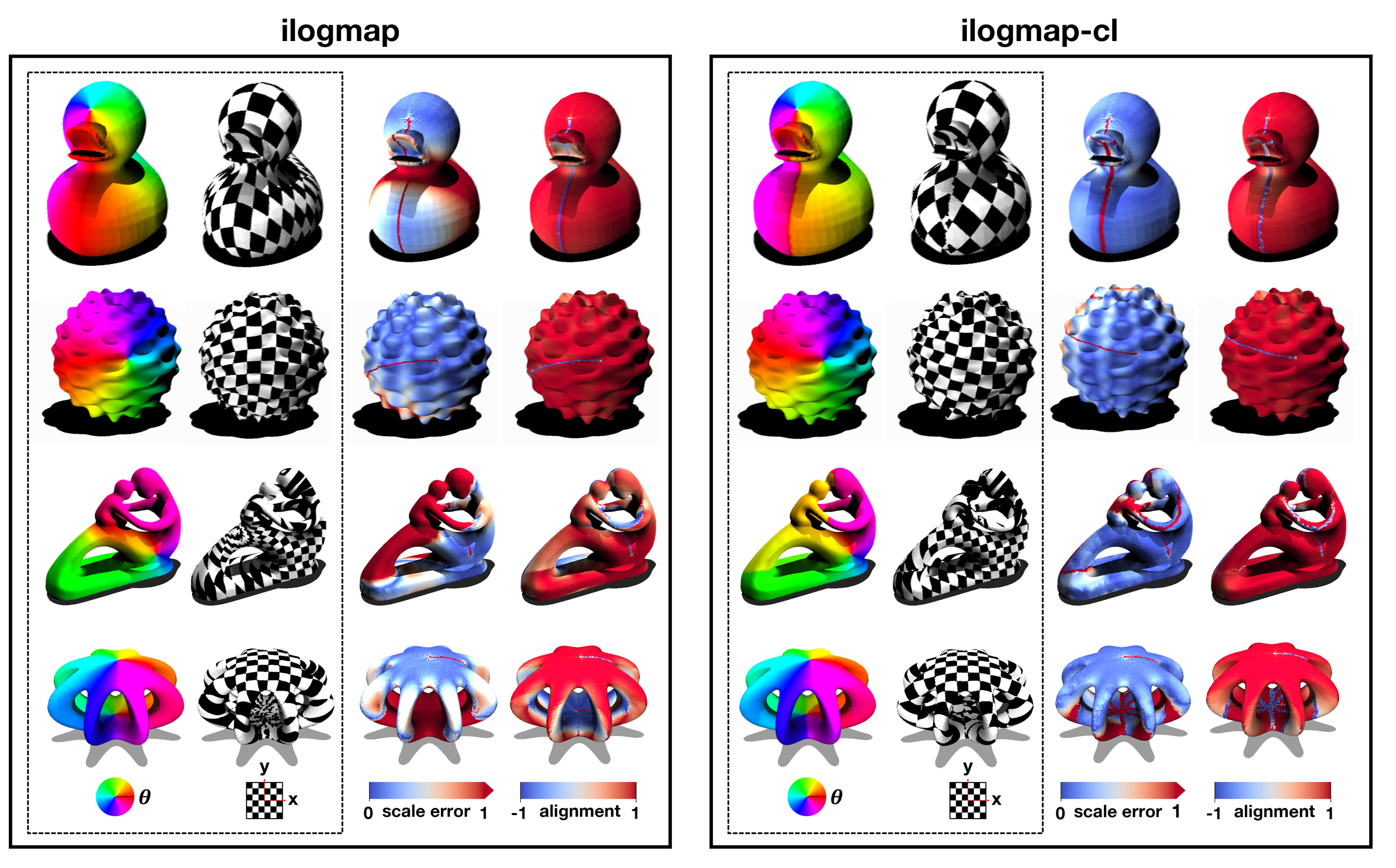}
\caption{iLogMap without (ilogmap, left) and with (ilogmap-cl, right) cut locus removal on a duck mesh, a bumpy sphere, a fertility model and a spider cage model. Each block shows the angular field $\theta$, a cartesian checkerboard texture induced by the GPC parameterization, the pointwise scale error and the alignment with respect to the magnetic vector field.
}
\Description{A gallery of four shapes (a duck, a bumpy sphere, a fertility statue model, and a spider cage model), each shown with the iLogMap and iLogMap-cl angular field, a cartesian checkerboard texture, a pointwise error between the Jacobi scale factor and the norm of the gradient of the computed angle, and a pointwise alignment measure between the circumferential direction and the gradient of the computed angle; iLogMap-cl displays less warping in the checkerboard texture, lower scale error and increased alignment values away from the source.}
\label{fig:gallery-cl}
\end{figure*}

We evaluate the two variants of the iLogMap method for a collection of closed surfaces with varying genus and increasing geometric complexity. Predictions are shown side by side in Figure~\ref{fig:gallery-cl}, and display four measurements: the angular field $\theta$ alongside its associated cartesian checkerboard texture, the scale approximation error and the angular alignment with respect to the connection field input. The scale error corresponds to the relative absolute difference $|1-h/\tilde{h}|$ between the predicted scale factor $\tilde{h}:=1/||\nabla\theta||$ and the solution $h$ to the Jacobi equation~\eqref{eq:jacobi-transport} obtained with FIM. The angular alignment is defined by the cosine of the angle formed between the gradient $\nabla\theta$ of the predicted phase $\theta$ and $\hat{\mathbf{e}}_\theta$. 

For all cases, both variants of iLogMap reconstruct uniform and well-structured angular phases from the prescribed connection, although their global behavior differs substantially. While iLogMap produces accurate estimates of the polar angle in the vicinity of the source point, it effectively collapses the cut locus into additional phase singularities. This smooths out discontinuities but, in turn, introduces distortion in the regions where these singularities appear. Such distortion is evidenced by the warping of the cartesian texture map, changes in the local scale factor, and increasing misalignment at locations farther from the source.

In contrast, removing the cut locus from the computation of the polar angle significantly improves the preservation of the cartesian pattern structure. The remaining errors are largely confined to the vicinity of the cut locus, where moderate scale distortion and local misalignment persist.

\subsection{Heterogeneous metrics}
To demonstrate the capacity of our method to generate anisotropic and heterogeneous GPCs, we evaluate it on a flat-plate model with unitary diameter under two regimes: a heterogeneous isotropic regime and a uniformly anisotropic regime.
\begin{figure}[t]
\centering
\includegraphics[width=0.45\textwidth]{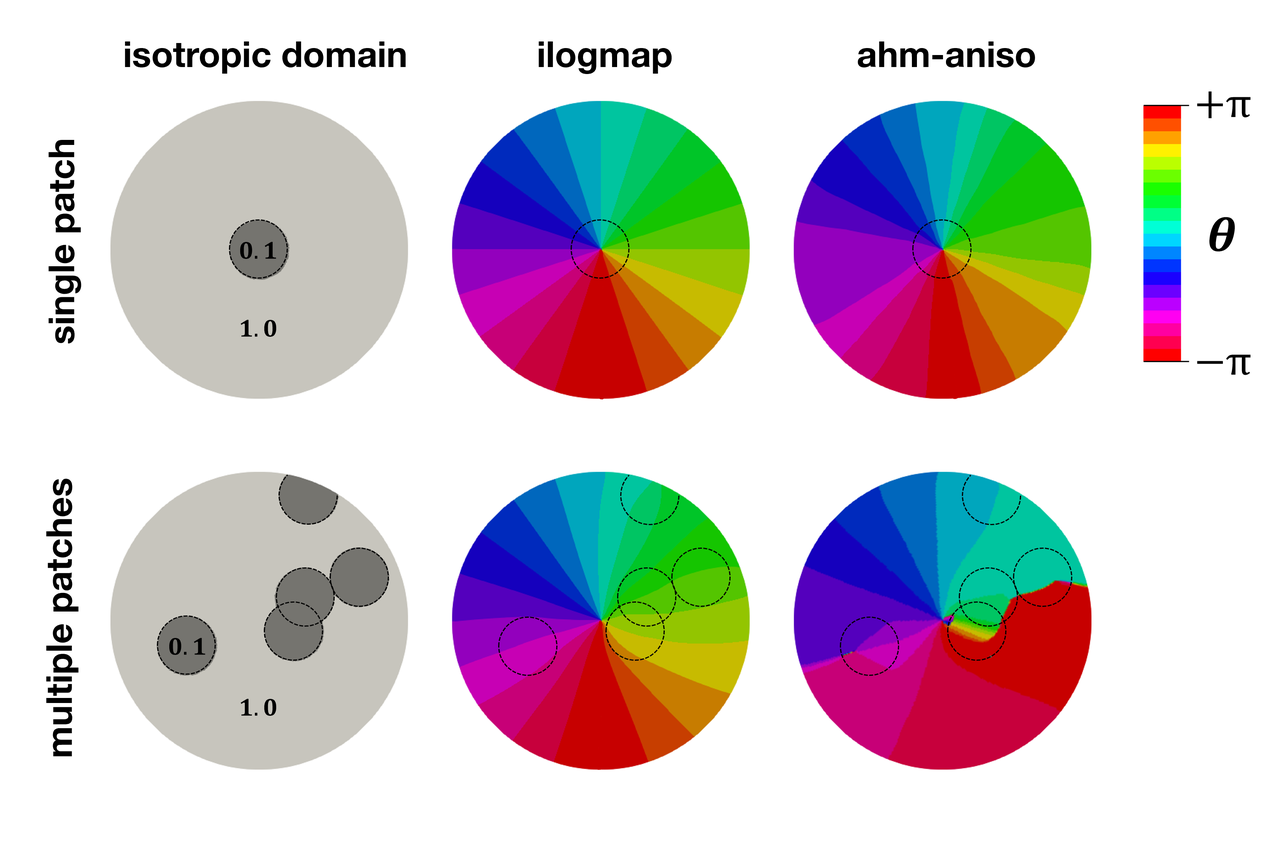}
\caption{Comparison of angle prediction from iLogMap and AHM-aniso for heterogeneous isotropic conductivity over a flat plate. We test two configurations: (i) A single circular center patch with low conductivity (first column, top) and (ii) five circular random patches with low conductivity (first column, bottom). Inside each patch, the geodesic distance is dilated by $1/\!\sqrt{\sigma}$.}
\Description{Rendering comparison between iLogMap and the anisotropic affine heat method for a flat circular plate under a heterogeneous isotropic conductivity regime: top row shows a single circular low-conductivity patch at the center of the domain, bottom row shows five randomly placed low-conductivity patches, with the angular level sets deflecting inside of each patch.}
\label{fig:iso}
\end{figure}

\begin{figure}[t]
\centering
\includegraphics[width=0.45\textwidth]{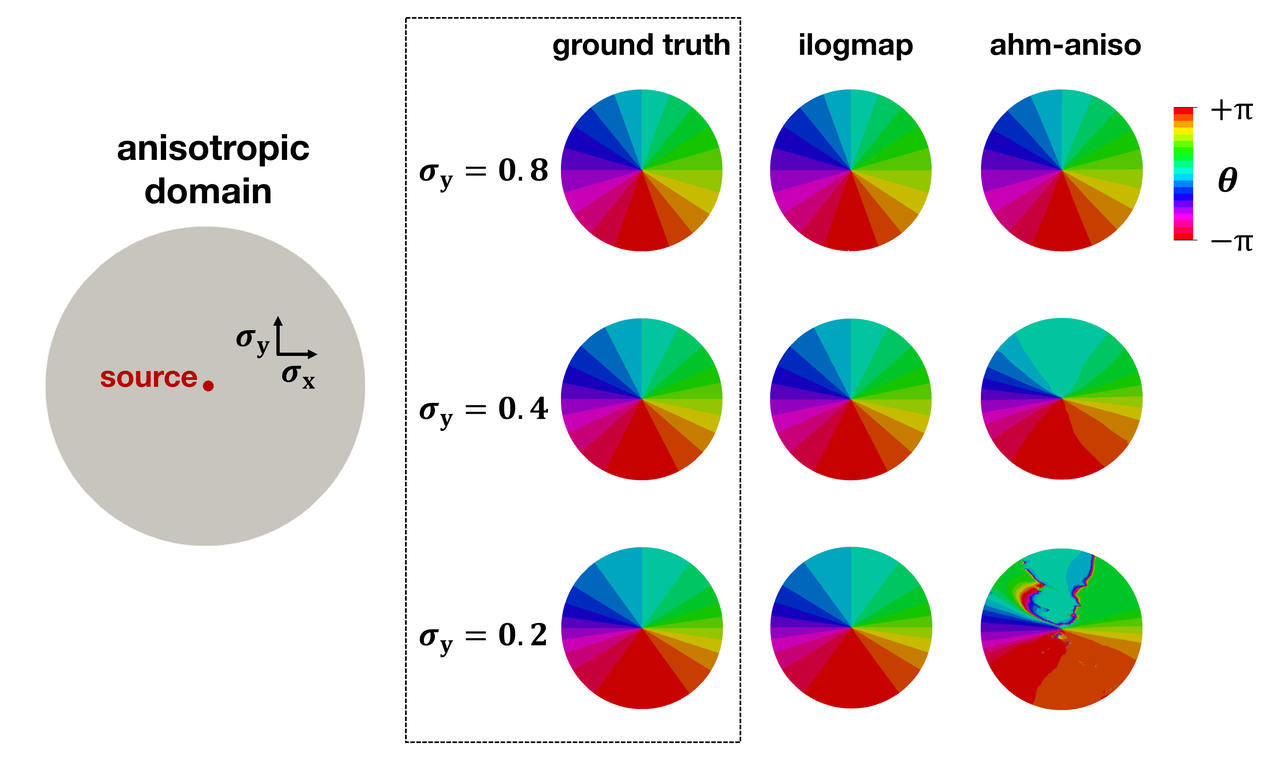}
\caption{Comparison of angle prediction from iLogMap and AHM-aniso for homogeneous anisotropic conductivity over a flat plate with fixed conductivity on the horizontal direction and decreasing conductivity on the vertical direction. The ground-truth angle is $\mathrm{arctan2}\bigl(y/\!\sqrt{\sigma_y},\,x/\!\sqrt{\sigma_x}\bigl)$.}
\Description{Angular outputs from iLogMap and the anisotropic affine heat method compared with the analytic ground truth angular field of a flat circle under a uniformly anisotropic conductivity regime with fixed horizontal conductivity and decreasing vertical conductivity. Angular bands accumulate toward the horizontal axis as vertical conductivity becomes smaller; iLogMap reproduces this behavior without distortion near the boundary, while the solution from AHM-aniso degrades noticeably.}
\label{fig:aniso}
\end{figure}

\subsubsection{Heterogeneous Isotropic Regime} 
We consider an heterogeneous medium with an isotropic metric $\sigma$ defined by the scalar field
\[
    \sigma := \begin{cases} 0.1, & \mathbf{x} \in P,\\[4pt] 1.0, & \mathbf{x} \notin P, \end{cases}
\]
where $P$ represents a region inside of the domain. The associated metric is then defined as $G(\mathrm{x}):=\sigma (\mathrm{x})\mathrm{id}$.

We test two configurations for the $P$ region:
\begin{enumerate}
\item A single circular inclusion of radius 0.1 centered on the plate.
\item A union of randomly sampled circular inclusions, each of radius 0.1.
\end{enumerate}  
We present both cases in Figure~\ref{fig:iso} and show the phase estimates for iLogMap and AHM-aniso. For the single circular inclusion (top), the angular component remains unaffected by the presence of the heterogeneity, presenting AHM-aniso a slight distortion of the phase towards the boundary. For the randomly distributed patches, our method deflects the phase map smoothly at the sectors with heterogeneous inclusions. In contrast, AHM with heterogeneous coefficients presents accentuated deflections and unsought phase jumps of the polar angle at the patch regions.

\subsubsection{Uniform Anisotropic Regime} In this setting, we prescribe a preferred direction for geodesic propagation along a given axis. This behavior is modeled by a diagonal metric tensor $G = \mathrm{diag}(\sigma_x, \sigma_y)$. The analytical form of the angular coordinate associated with this metric is $\theta = \mathrm{arctan2}(y/\!\sqrt{\sigma_y},\,x/\!\sqrt{\sigma_x})$.

When $\sigma_y<\sigma_x$, geodesics stretch along the horizontal direction, causing the level sets of the angular phase to concentrate toward the horizontal axis. This phenomenon is accurately reproduced by our method in Figure~\ref{fig:aniso}, which shows the estimates of iLogMap and AHM-aniso alongside the ground truth values for the angular coordinate. AHM-aniso produces comparable estimates only when the anisotropy ratio $\sigma_x:\sigma_y$ is close to unity. As this ratio increases, the method exhibits progressive deterioration and ultimately fails to recover an accurate log-map parameterization under strong anisotropy.

\subsection{Volumetric Coordinates}\label{sec:vol-results}

\begin{figure*}[t]
\centering
\includegraphics[width=\textwidth]{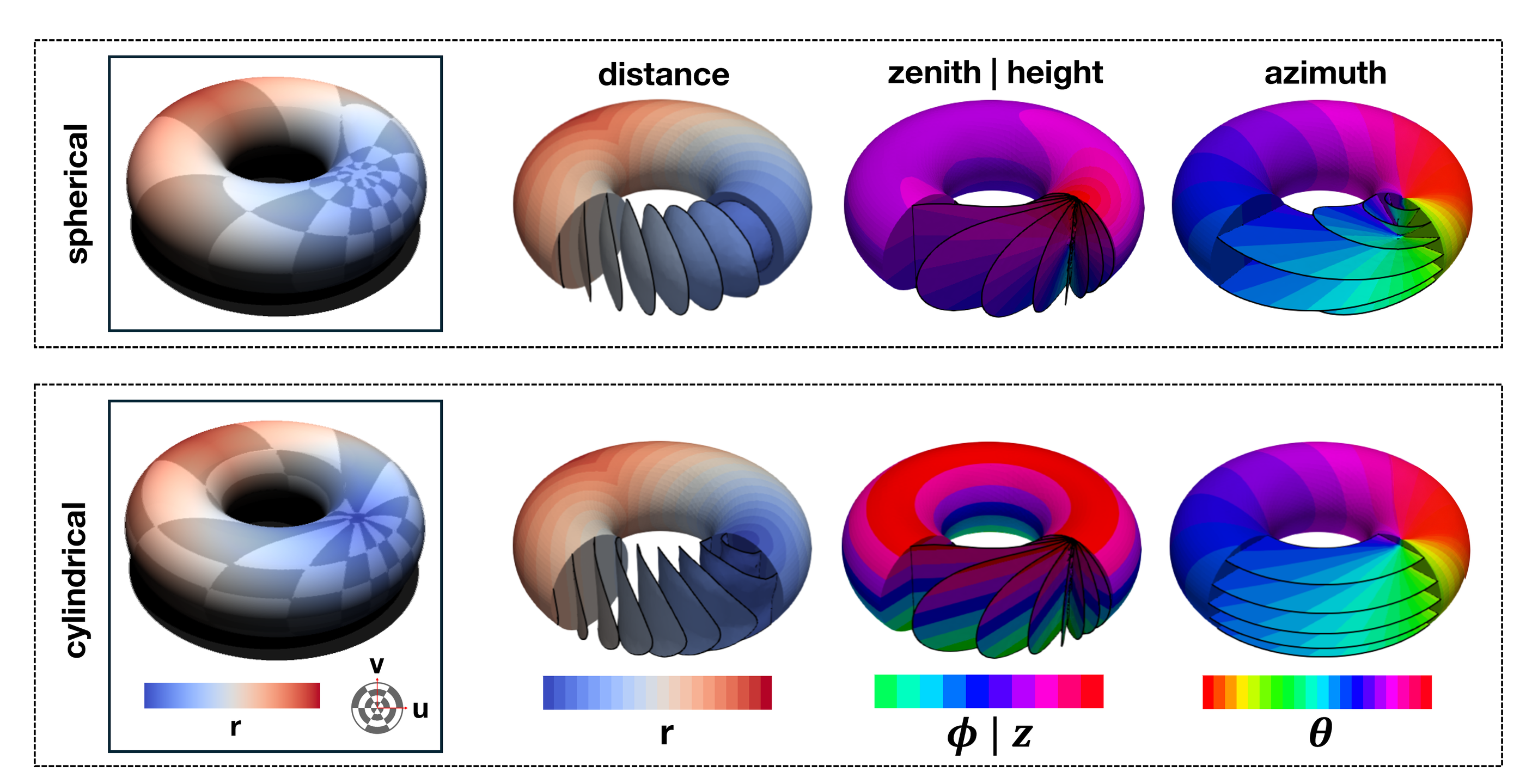}
\caption{Spherical (top) and cylindrical (bottom) volumetric parameterizations applied to a solid torus. We display each coordinate with a peeled section of the volume to illustrate the variation of each component away from the source.}
\Description{A solid torus mesh with peeled cross sections according to spherical/cylindrical parameterizations, showing the spherical volumetric coordinate field on top and the cylindrical volumetric coordinate field on the bottom.}
\label{fig:torus-vol}
\end{figure*}

Figure~\ref{fig:spherical} shows iLogMap applied to solid tetrahedral meshes of a unit sphere and a cylinder (height = 2.0, radius = 1.0) with maximum edge length resolution of 0.1. On the sphere, the spherical parameterization~\eqref{eq:sph-ephi} yields an estimate close to the expected spherical fields. We report relative $L_1$ errors of 3.4\% for the geodesic distance, 2.9\% for the complex residual of the zenith angle and 2.4\% for the complex residual of the azimuthal angle. For the case of the cylinder (Figure~\ref{fig:cylinder}), a similar behavior is observed, with a relative $L_1$ error of 3.3\% for the radial distance, 1.4\% for the axial coordinate and 0.8\% for the complex residual of the polar angle.

Figure~\ref{fig:torus-vol} shows both coordinate systems applied to a solid torus, confirming that the parameterization is selectable independently of mesh topology.
%The larger errors on the cylindrical coordinates approximation reflect accumulated error from an artificial oscillation of the filament singularity caused by mesh resolution. 
\subsection{Applications in Computational Cardiology}

Over the past decade, computational models have emerged as powerful tools for investigating complex biological systems and their multiscale interactions. In cardiovascular medicine, patient-specific computational models have attracted particular interest due to their ability to reconstruct cardiac electrophysiology, non-invasively assess disease mechanisms and evaluate therapeutic strategies before clinical intervention \cite{trayanova2011whole, niederer2011, kim2019}. By integrating medical imaging and electrophysiological measurements, these digital representations of the heart can simulate the response of individualized patients to therapy, extract tissue properties and provide mechanistic insights of cardiac function \cite{grandits2025,emagana,mcdowell2015,vigmond2008}.

In this section, we demonstrate two applications of our method in computational cardiology: (i) the generation of spiral phase maps for the initialization of synchronized arrhythmic events on atrial surfaces and (ii) the construction of eikonal solutions in volumetric ventricular domains.

\subsubsection{Spiral Phase Initialization}

Geodesic polar coordinates provide a natural basis for extending closed-form solutions to the complex eikonal equation from flat Euclidean space to curved surfaces. By employing the GPC parameterization on the atrial geometry from a source point, logarithmic maps can be used to construct spiral phase maps that capture the intrinsic geometry of the cardiac domain. The phase singularity distribution serves as an initiator of stable re-entrant patterns of tissue excitation, allowing to control rotor dynamics in simulation that are frequently associated with atrial fibrillation \cite{azzolin2021peerp,jacquemet2010,jacquemet2012,bezerra2025}.
\begin{figure}[t]
\centering
\includegraphics[width=0.4\textwidth]{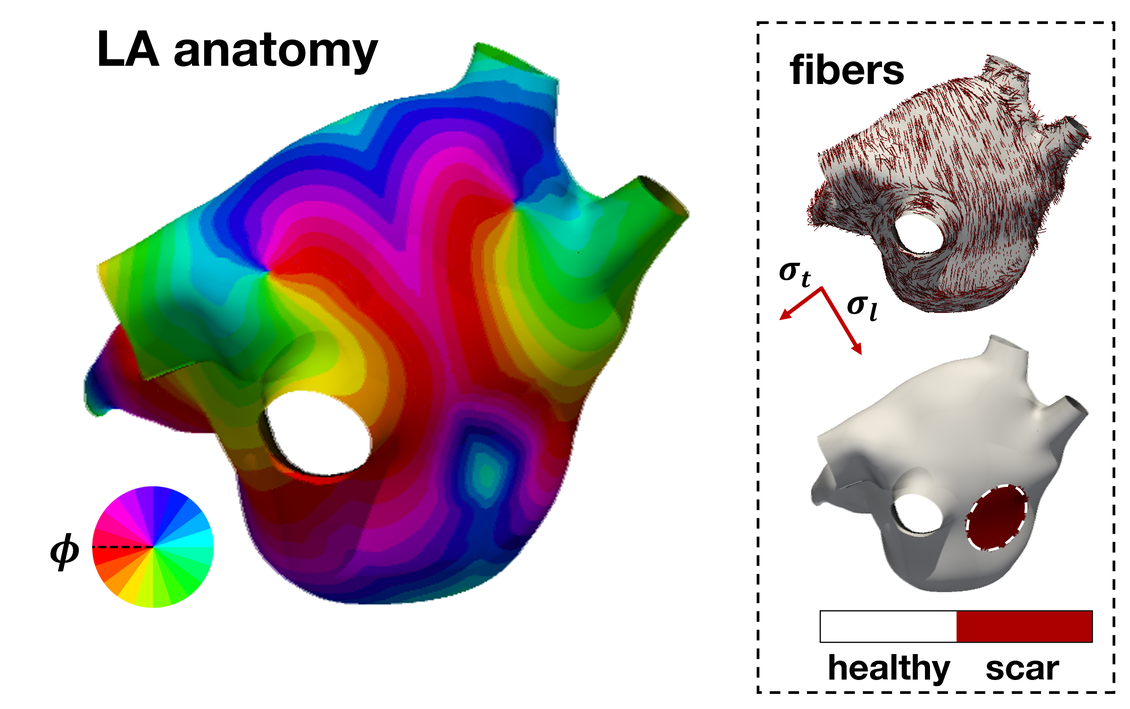}
\caption{Example of synchronized spiral phases in the presence of fibers ($\sigma_t:\sigma_l=1:3$) and scar tissue ($\sigma_\text{scar}:\sigma_l=1:9$) for an atrial model.}
\Description{Left: A left atrial surface mesh colored with a phase field phi from two synchronized spirals for a heterogeneous anisotropic conductivity regime using iLogMap. Anisotropy is induced by cardiac fibers and scar. Right: top, distribution of fibers; bottom, single scar tissue patch.}
\label{fig:LA-spirals}
\end{figure}
This approach can be further generalized to create multiple interacting sources on the atrial surface \cite{banduc2025}. However, when several sources are created simultaneously, the corresponding phase fields must be synchronized across their interface to prevent discontinuities in the phase distribution. Such discontinuities may introduce artificial phase singularities that lead to the formation of spurious spiral waves, thus compromising the physiological relevance of the simulation.

The global angular parameterization produced by iLogMap naturally resolves this limitation. Since our method constructs a globally consistent phase field, multiple spiral sources can be initialized directly without requiring posterior alignment (see Figure~\ref{fig:LA-spirals}). This property facilitates the generation of complex, synchronized arrhythmic activation patterns while preserving phase continuity on the entire cardiac surface.

\subsubsection{Eikonal Solution Generation}\label{sec:cardiac-results}
Computational models of cardiac electrophysiology typically rely on high spatial resolution for the accurate representation of myocardial excitation, making them computationally demanding in many practical applications \cite{franzone2014,vigmond2008,trayanova2011whole}. A widely used approximation for fronts of electrical activation is the eikonal equation, which provides an efficient way of computing local activation times in anisotropic cardiac tissue by reducing the monodomain model to a normal wavefront while preserving its propagation characteristics \cite{franzone2014, Pezzuto2017Fast}.

The extended version of iLogMap enables the construction of generalized cylindrical GPC parameterizations for volumetric ventricular geometries incorporating myocardial fiber architecture. By shifting the coordinate sources, our method can generate eikonal activation patterns by computing Euclidean distances in the GPC space in a computationally efficient and geometrically consistent manner. 

\begin{figure}[t]
\centering
\includegraphics[width=0.4\textwidth]{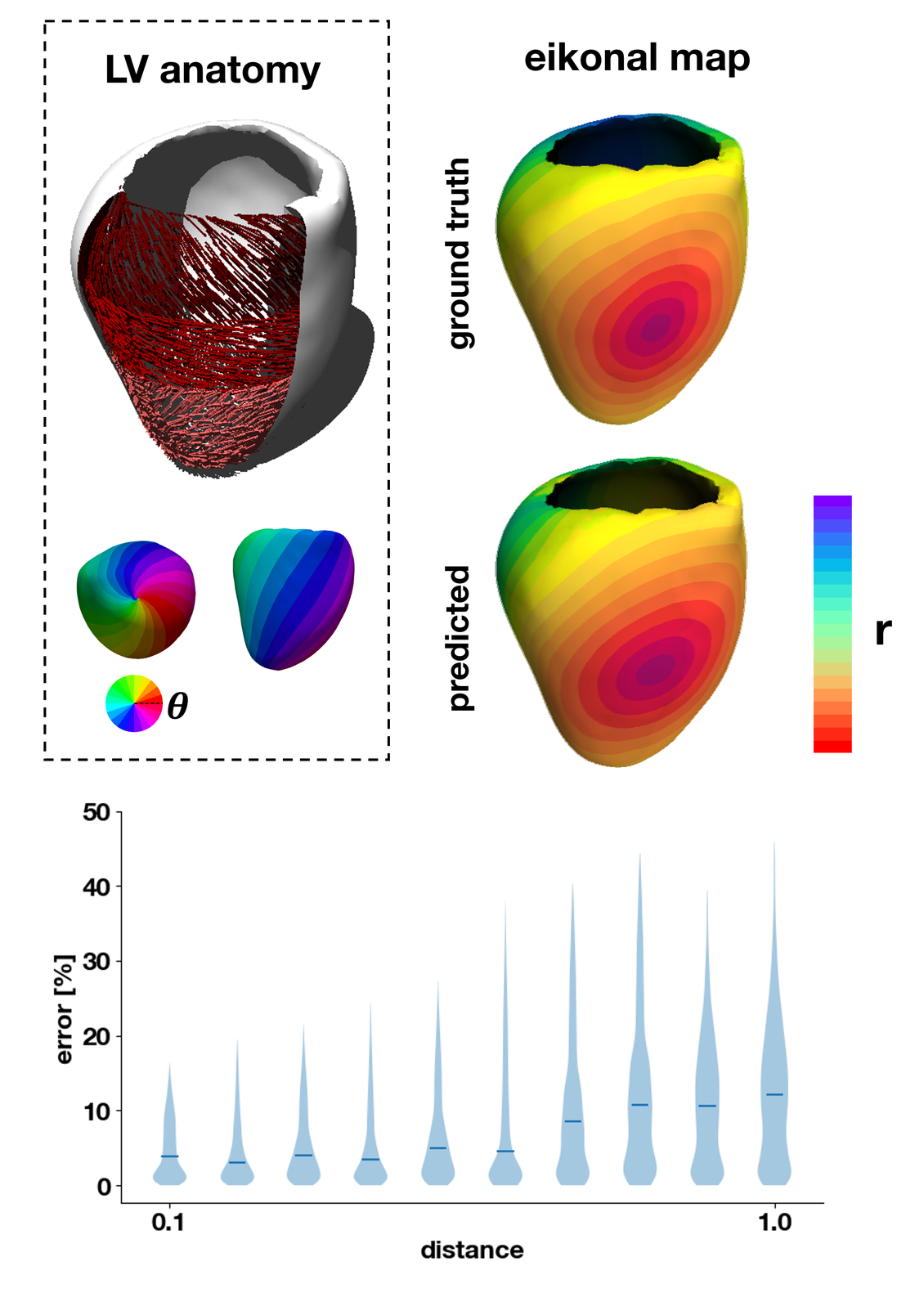}
\caption{iLogMap applied to a ventricular model. Top: Left ventricular geometry with fibers ($\sigma_t:\sigma_l=1:2$). We show the angular component from an apical base point (left) and the resulting activation pattern from local shifting, together with a ground truth eikonal solution (right). Bottom: Relative geodesic distance error for increasing source-point displacement. The error is computed as the absolute difference between the ground truth and iLogMap solutions, normalized by the maximum real distance value.}
\Description{Top: left, a left ventricular volume mesh with cardiac fibers, shown through a transmural peeling, and views of the angular component from cylindrical coordinates using iLogMap from a base point at the apex under a uniformly anisotropic conductivity regime; the angle component presents torsion following fiber distribution; right, ground truth and predicted eikonal maps using base-point shifting over the resulting parameterization, with visually similar activation patterns. Bottom: violin plots of pointwise absolute difference between predicted activation maps and ground truth activation maps, normalized by maximum geodesic distance, for increasing base-point shift distance; median errors increase as points farther away from the original source are considered.}
\label{fig:lv-eikonal}
\end{figure}

In Figure~\ref{fig:lv-eikonal} we show a left ventricular model \cite{deMarvao2014, Bai2015} with synthetic cardiac fibers that introduce torsion to the angle component. The base point is initially located at the apex and perturbed to generate eikonal estimates. For each shifted source, we compute the new distance over the entire geometry and compare it with a baseline geodesic solution obtained with FIM \cite{grandits2021}. As the source is displaced farther from the apex, error increases progressively, with median relative errors ranging from 3.0\% to 12.1\%.

\section{Discussion}

iLogMap is a FEM framework that reformulates the angular component of the logarithmic map computation as an angular synchronization problem lifted to a complex directional field. This problem is solved globally via an eigenproblem of a magnetic Laplacian constructed from the rotational field derived from the geodesics emanating from a base point. This perspective offers several conceptual and practical advantages over existing methods.

\textbf{Problem Decoupling.} By decoupling the radial coordinate from the polar angle reconstruction, iLogMap allows independent control over each GPC component. Different geodesic distance solvers can be adopted without modifying the angular synchronization step and the accuracy of each component can be tuned separately. In contrast, the affine heat method computes both components simultaneously through affine diffusion, making it harder to separately improve the parameterization in some cases, particularly in domains with bottleneck topology, where short-time asymptotics degrade into distorted geodesics propagation, and surfaces with boundaries, where the diffusion model can suffer from reflection artifacts \cite{crane2013heat,avramidi1995}.

\textbf{Jacobi Scale Factor.} The use of the Jacobi field $h$ as a weight coefficient is geometrically motivated: it precisely quantifies the angular spread of geodesics under surface curvature. The formulation accommodates simpler scale factor estimates and can modify in a straightforward way how geodesics vary radially. For instance, $h = r$ is exact on flat domains and may suffice on mildly curved surfaces (see Figure~\ref{fig:scale_factor_comparison}). However, solving equation~\eqref{eq:jacobi-transport} is recommended for general curved geometries. For solvers that do not expose individual update steps, the advection--diffusion system~\eqref{eq:jacobi-transport2} provides a reliable alternative.

\textbf{Robustness Near the Cut Locus.} The cut locus presents an inherent topological challenge to constructing a continuous polar angle field. Across this region, the angular coordinate becomes multi-valued, leading to phase discontinuities that may be undesirable in applications requiring low distortion of the logarithmic map. iLogMap mitigates this issue by optionally removing the cut locus prior to solving the synchronization eigenproblem (Algorithm~\ref{algo:ilogmap-cl}) and subsequently extending the field harmonically. Our experiments demonstrate that this strategy consistently decreases distortion at fold regions. Alternatively, disabling the cut locus removal allows the discontinuity to collapse into isolated phase singularities, yielding a more regular angular field at the expense of increased local distortion around these points.

\textbf{Anisotropic Conditions and Volumetric Extension.} Our workflow can be easily extended to strong anisotropic conditions and can be adapted to volumetric domains.

\textbf{Limitations.} The finite element matrix $\mathbf{L}_\theta$ must be recomputed whenever the source point changes, since the circumferential field $\frac{1}{h}\hat{\mathbf{e}}_\theta$ depends explicitly on the source location. The adaptive variant of the AHM employed in this work shares this limitation, requiring a new factorization for each base point. In contrast, the localized variant of the AHM \cite{soliman2025affine} admits a reusable factorization across different sources. A similarly pre-computable formulation for iLogMap would therefore be highly desirable for applications involving many source points separately.

Another limitation arises from the cut locus detection procedure (Algorithm~\ref{algo:ilogmap-cl}), which requires selecting a threshold parameter $t$ (we fixed the value $t=\pi/4$). Although the method is relatively insensitive to the choice of $t$ over a broad range of values, an automatic selection strategy remains an open problem \cite{mancinelli2021practical}, and one from which our method would directly benefit.

Finally, our formulation has been validated only for two-di\-men\-sio\-nal manifolds and for three-dimensional volumetric domains. Extending this approach to higher-dimensional manifolds is non-trivial. In particular, each additional intrinsic dimension requires computing an extra coordinate field, while the associated angular synchronization problem becomes more complex. Moreover, the corresponding Jacobi equations require shape operators whose interactions may become increasingly coupled and difficult to characterize \cite{doCarmo1992}. Developing a general framework for higher-dimensional parameterizations is an important direction for future work.

\section{Conclusion}

We introduced iLogMap, a novel method for computing geodesic polar coordinate parameterizations on surface meshes. The premise of our approach is that recovering the angular component of the logarithmic map is equivalent to solving an angular synchronization problem over a circumferential vector field, which admits a clean spectral relaxation as the ground-state eigenfunction of a magnetic Laplace operator. Combined with an enhanced fast iterative method \cite{grandits2021} that computes both the geodesic distance and an auxiliary Jacobi scale factor, iLogMap produces accurate and globally consistent polar coordinate parameterizations with a straightforward finite element implementation. 

Experiments on canonical shapes with known parameterizations confirm convergence with mesh refinement. Comparisons against the affine heat method and a naive Laplace interpolator on tori with varying genus, geometries with boundary and several synthetic shapes show that iLogMap with the cut locus removal achieves competitive angular accuracy and decreased metric distortion, particularly on surfaces with intricate topology or high curvature. The parameterization induced by our method is of sufficient quality to be used in practical applications \cite{soliman2025affine}.

Experiments on flat-plate benchmarks with heterogeneous iso\-tro\-pic patches and uniform anisotropic conductivity demonstrate that the anisotropic extension accurately reproduces the expected elliptical level sets and sheared coordinate lines, while outperforming the anisotropic variant of the affine heat method baseline in terms of angular accuracy. This capability is directly relevant to computational cardiology, where our method can be employed to initiate synchronized spiral phases in an anisotropic atrial model \cite{Potse2015} with fibrosis using a single angular synchronization solve \cite{banduc2025}.

The volumetric extension of iLogMap (Section~\ref{sec:volumes}) demonstrates that the same eigenproblem formulation applies to solid tetrahedral meshes, recovering spherical and cylindrical parameterizations in a torus by replacing the surface normal with an externally supplied axial direction and extending the source singularity from a point to a filament. The cardiac application (Section~\ref{sec:cardiac-results}) shows that anisotropic iLogMap coordinates on a left ventricular mesh \cite{deMarvao2014, Bai2015} follow fiber rotation and can produce eikonal estimates at locations away from the base point.

We believe that our method is a contribution to the geometry processing toolbox and expands the capability of current parameterization techniques. %A repository containing the core implementation of our method and a simple geometric benchmark is available at \hyperlink{ https://github.com/HiddenHeartLab/ilogmap-demo.git}{https://github.com/HiddenHeartLab/ilogmap-demo.git}.

%\begin{acks}
%TB and FSC acknowledge the support of the Millennium Institute for Intelligent Healthcare Engineering i-Health, ICN2021\_004. TB and FSC also acknowledge the financial support of the project ERAPERMED-134 from ANID. SP acknowledges the support of the SNF\-FWF ``CardioTwin'' project (no.~214817), the PRIN-PNRR project no.~P2022N5ZNP, CSCS production grant no.~s1275, and INdAM-GNCS. TB, SP and FSC acknowledge the guidance of Nicholas Sharp to adapt his code for the anisotropic variant of the logarithmic map.
%\end{acks}
\bibliographystyle{ACM-Reference-Format}
\bibliography{bibliography}

%\appendix

\end{document}